\documentclass[aps,prl,twocolumn,a4paper,showpacs,10pt]{revtex4-2}
\usepackage{graphicx}
\usepackage{amsmath}
\usepackage{amssymb}
\usepackage{mathrsfs}
\usepackage{amsfonts}
\usepackage{dsfont}
\usepackage{dcolumn}% Align table columns on decimal point
\usepackage{bm}% bold math
\usepackage{booktabs}
\usepackage{color}
\usepackage{xcolor}
\usepackage{colortbl}
\usepackage{mathtools}
\usepackage{upgreek}
\usepackage{cancel}
\newcommand{\inp}{\mathrm{in}}
\newcommand{\out}{\mathrm{out}}
\newcommand{\as}{\mathrm{as}}

\newcommand{\ha}{\hat{a}}
\newcommand{\had}{\hat{a}^\dagger}
\newcommand{\hb}{\hat{b}}
\newcommand{\hbd}{\hat{b}^\dagger}

\newcommand{\hPi}{\hat{\Pi}}
\newcommand{\hPsi}{\hat{\Psi}}

\newcommand{\hpm}{\hat{A}^{\dagger}}
\newcommand{\hpp}{\hat{A}}

\newcommand{\hS}{\hat{S}}

\newcommand{\hH}{\hat{H}}

\newcommand{\hN}{\hat{N}}

\newcommand{\mx}{\mathbf{x}}

\newcommand{\mr}{\mathbf{r}}

\newcommand{\du}{{(2)}}

\newcommand{\di}{\mathrm{d}}
\providecommand{\abs}[1]{\left|#1\right|}

\providecommand{\ket}[1]{|#1\rangle}
\providecommand{\bra}[1]{\langle#1|}
\providecommand{\brak}[2]{\langle#1|#2\rangle} % si usa come: \brak{A}{B} = <A|B>
\providecommand{\mean}[3]{\langle#1|#2|#3\rangle} % si usa come: \mean{A}{B}{C} = <A|B|C>

\begin{document}

\title{A Simple Field Theoretic Description of Single-Photon Nonlocality}

\author{Andrea Aiello}
\email{andrea.aiello@mpl.mpg.de} % optional
\affiliation{Max Planck Institute for the Science of Light, Staudtstrasse 2, 91058
Erlangen, Germany}
% Please provide a full mailing address here.

% See the REVTeX documentation for more examples of author and affiliation lists.

\date{\today}

\begin{abstract}
We present a simple yet rigorous field theoretic demonstration of the nonlocality of a single-photon field. The formalism used allows us to calculate the electric field of a single-photon light beam sent through a beam splitter, which directly demonstrates that it is the light field, rather than the photon itself regarded as a particle, that exhibits nonlocality. Our results are obtained  without using either inequalities or specific measurement apparatuses, so that they  have  perfectly general validity.
\end{abstract}

\maketitle

%%%%%%%%%%%%%%%%%%%%%%%%
\emph{Introduction.--}
%%%%%%%%%%%%%%%%%%%%%%%%
In 1986 Grangier, Roger and Aspect published a seminal paper \cite{Grangier_1986} reporting about
two experiments where a light beam prepared in a single-photon quantum state was sent through  a beam splitter. In the first experiment two photomultipliers were placed behind the output ports of the beam splitter. As predicted by quantum mechanics, Grangier \textit{et al.},  found that per each run of the experiment only one of the two detectors could fire. In the second experiment, a Mach-Zehnder interferometer was built by coupling the two output ports of the original beam splitter to the input ports of a second beam splitter, at the output ports of which  the two photomultipliers were now settled.
With this setup Grangier and coworkers could observe interference fringes by changing the path difference between the two arms of the interferometer.

According to quantum mechanics the light after the first beam splitter in both the above experiments can be described by the state vector $\ket{\psi}$ defined by \cite{LoudonBook,PhysRevA.72.064306,PhysRevA.74.026301,PhysRevA.74.026302},
\begin{align}\label{a5}
\ket{\psi}  = \frac{1}{\sqrt{2}} \bigl( \ket{1}_1 \ket{0}_2 + i \ket{0}_1 \ket{1}_2 \bigr)  ,
\end{align}
where the subscripts $1$ and $2$ label the two modes associated with the two  ports of the beam splitter. If we insist for a particle description of the state \eqref{a5}, we must maintain that the photon exits both ports of the first beam splitter,  a phenomenon often referred to as single-photon nonlocality  \cite{PhysRevLett.66.252,PhysRevLett.74.4571,
PhysRevLett.75.2064,PhysRevLett.73.2279,PhysRevA.64.042106,PhysRevLett.92.180401,Drezet2006,PhysRevLett.99.180404,PhysRevA.84.012110,Fuwa2015}.
The concepts of single-photon states and single-photon nonlocality, other than stimulating an ongoing lively debate  \cite{PhysRevA.73.042111,Englert2013,leuchs2015physical}, have proven to be extremely useful for many quantum communication and quantum computation applications as, for example, the implementation of universal quantum gates \cite{PhysRevA.57.R1477,Knill2001}, teleportation \cite{PhysRevLett.88.070402}, entanglement swapping \cite{PhysRevA.66.024309}, and many others purposes \cite{PhysRevA.35.2532,PhysRevA.63.012305,Choi2008,PhysRevLett.104.180504}.
In this paper we will find a unique quantum field theory representation of the single-photon state \eqref{a5} that, differently from all previous works, allows a demonstration of nonlocality without using either inequalities or specific measurement apparatuses.
 Our main result is that the physically observable electric field of light in the  single-photon state  \eqref{a5}, measured simultaneously at two different locations behind the two ports of the beam splitter, is completely determined by the  field associated with the single-photon state entering the device.
The use of quantum field theory to explain optical interference phenomena is becoming increasingly popular nowadays \cite{doi:10.1119/1.10857,PhysRevD.104.065013,PhysRevLett.125.040401,PhysRevResearch.2.033266,vedral2021quantum}.
\\
%
%%%%%%%%%%%%%%%%%%%%%%%%
\emph{Theory.--}
%%%%%%%%%%%%%%%%%%%%%%%%
%
Many experiments in optics use  monochromatic collimated light beams with uniform polarization. The electric
field of such beams can be expressed in terms of any complete set of basis functions of the form $u(\mx,z) = \varphi(\mx,z) \exp(i kz)$, where $\varphi(\mx,z) $ denotes a  solution  of the paraxial wave equation: $\left( \partial^2 / \partial x^2 +
\partial^2 / \partial y^2 + 2 i k \partial / \partial z \right) \varphi(\mx,z) = 0$, with $z$ being the direction of
propagation and $k = \omega/c = 2 \pi / \lambda$  the wavenumber of light of frequency $\omega$ and wavelength
$\lambda$ in vacuum. Here and hereafter $c$ is the speed of light in vacuum, and $\mx = x \, \bm{e}_x + y \, \bm{e}_y$
is the transverse position vector on the $xy$-plane perpendicular to the axis $z$.
Typical basis functions are the two-dimensional Hermite-Gauss  modes, denoted $u_{\mu}(\mx,z) = \varphi_{n_\mu}(x,z)
\varphi_{m_\mu}(y,z)\exp(i kz) = \varphi_{\mu}(\mx, z) \exp(i kz)$, where  $n_\mu,m_\mu=0,1,2, \ldots, \infty$. Throughout this paper, Greek letters $\mu, \mu', \dots$, denote distinct ordered pairs of nonnegative integer indexes: $\mu = (n_\mu,m_\mu), \; \mu' = (n_{\mu'},m_{\mu'})$, etc. Hermite-Gauss modes $u_{\mu}(\mx,z)$ are also commonly referred to as $\text{TEM}_{n_\mu m_\mu}$ modes. The one-dimensional Hermite-Gauss modes are  defined by
\begin{align}\label{a10}
\varphi_{n}(x,z) = & \; \frac{1}{\pi^{1/4}} \frac{1}{ \sqrt{2^n n!}} \, \frac{1}{\sqrt{x_0}} \, \text{H}_n \! \left( { x}/{x_0} \right) e^{-(x/x_0)^2/2} \nonumber \\[4pt]
& \times  e^{\frac{i}{2}\frac{z}{z_0} (x/x_0)^2} \, e^{ - i \left( n + \frac{1}{2} \right) \arctan \left( {z}/{z_0} \right) },
\end{align}
$(n= 0,1,\ldots, \infty)$, where $\text{H}_n(x)$ is the  $n$th-order Hermite polynomial, and
$x_0 =  w_0[(1+ z^2/z_0^2)/2]^{1/2}$
fixes the transverse length scale at distance $z$ from the beam's origin, where the minimum beam radius   $w_0 > 0$, is attained. The Rayleigh length $z_0 = \pi w_0^2/ \lambda$,  sets the longitudinal length scale, giving the distance over which the beam can propagate without spreading significantly \cite{Siegman}. For example,  a He-Ne laser beam with minimum radius $w_0 = 2 \; \text{mm}$, has $z_0 \simeq 20
\; \text{m}$, so that $w_0/z_0 \simeq 10^{-4} $ and the transverse and longitudinal degrees of freedom effectively decouple for $z \ll z_0$. This implies that across a table-top experimental setup with the linear size of about $1 \; \text{m}$, the radius
of such beam will vary by less than $0.125 \, \%$, so that it can be considered as practically constant. Therefore, in
the remainder the coordinate $z$ will be regarded as a constant parameter, as opposite to the dynamical variables $x$ and $y$, and we will write indifferently either $(\mx,z)$ or $(\mr)$, in the arguments of the functions.

The Hermite-Gauss modes form a complete and orthogonal set of basis functions on $\mathbb{R}^2$, i.e.,
\begin{align}
\int \di^2x \, u_{\mu}^*(\mx,z) u_{\mu'}(\mx,z) = & \;  \delta_{\mu \mu'}, \label{a20} \\[4pt]
\sum_{\mu} u_{\mu}(\mx,z)u_{\mu}^*(\mx',z) = & \; \delta^\du \left( \mx - \mx' \right) , \label{a30}
\end{align}
where $\delta_{\mu \mu'} = \delta_{n_\mu n_{\mu'}}\delta_{m_\mu m_{\mu'}}$, and $\delta^\du \left( \mx - \mx' \right)=\delta \left( x - x' \right)\delta \left( y - y' \right)$.  Here and hereafter
two-dimensional integrals are understood to be calculated over the whole $\mathbb{R}^2$ plane, and $\sum_{\mu}$   stands for $ \sum_{n_\mu=0}^\infty \sum_{m_\mu=0}^\infty$.
%
%\begin{align}\label{a40}
%
%\sum_{\mu} \qquad \text{stands for} \qquad \sum_{n_\mu=0}^\infty \sum_{m_\mu=0}^\infty.
%
%\end{align}
%

Consider now  a monochromatic  paraxial light beam propagating in the $z$ direction and polarized along the $x$ axis of a given Cartesian coordinate system.  In the Coulomb gauge, its electric field
will be well approximated by  $\mathbf{E}(\mr,t) = E(\mr,t) \, \bm{e}_x$, where
\begin{align}\label{a50}
E(\mr,t) =  \sum_{\mu} \left[ \alpha_{\mu}  e^{ -i \omega t} u_{\mu} (\mx,z) + \text{c.c.} \right],
\end{align}
with $\alpha_{\mu}$ denoting the time-independent complex amplitude of the field in the mode $u_{\mu} (\mx,z)$.
The simple form of Eq. \eqref{a50} enables us to proceed with a plain phenomenological quantization
akin to the familiar non-relativistic  second-quantization.
Thus, we replace the  amplitudes $\alpha_{\mu}$ and $\alpha_{\mu}^*$ in \eqref{a50} with the annihilation and creation operators $\ha_{\mu}$ and $\had_{\mu}$, respectively, which by definition satisfy the
bosonic canonical commutation relations
\begin{align}\label{a60}
\bigl[ \ha_{\mu}, \; \ha^\dagger_{\mu'} \bigr] = \delta_{\mu \mu'} .
\end{align}
Accordingly, the operator $ \ha_{\mu}$ annihilates a photon in the mode $u_{\mu} (\mx,z)$. With this substitution, we achieve $E(\mr,t) \to \hPsi (\mr,t)$, where
\begin{align}\label{a70}
\hPsi (\mr,t) =  \frac{1}{ \sqrt{2 \, \omega}} \left[ \hpp (\mx,z,t)  +  \hpm (\mx,z,t)  \right],
\end{align}
with
\begin{align}\label{a80}
\hpp (\mx,z ,t) = e^{- i \omega t} \, \sum_{\mu} \ha_{\mu} u_{\mu} (\mx,z) \,  .
\end{align}
The normalization prefactor $1/\sqrt{2 \, \omega}$ has been introduced for later
convenience.
%%%%%%%%%%%%%%%%%%%%%%%%%%%%%%%%%%%%%%%%%%%%%%%%%%%%%%%%%%%%%%%%%%%%%%%%%%%%%%%%%%%%%%%%%%%%%%%%%%%%%%%%%%%%%%%%%%%%%%%%%%
This elementary quantum field  model can be completed by introducing the conjugated  momentum operator $\hPi (\mx,z,t) = \di \hPsi (\mx,z,t)/ \di t$, such that
\begin{align}\label{a100}
\left[\hPsi (\mx,z,t)  , \, \hPi (\mx',z,t) \right] = i \, \delta^\du \left( \mx - \mx' \right).
\end{align}

Using (\ref{a20}-\ref{a30}), it is not difficult to show that the  time-independent  Hamiltonian
\begin{align}\label{a110}
\hH = & \; \frac{1}{2} \int \di^2x  \left[  \hPi^2 (\mx,z,t) + \omega^2 \hPsi^2 (\mx,z,t) \right] \nonumber \\[4pt]
   = & \; \omega  \sum_{\mu} \left( \had_{\mu} \ha_{\mu} + \frac{1}{2} \right),
\end{align}
generates the correct time-evolution for the field operators, that is $\hat{a}_{\mu}(t) = e^{ i \hH t} \, \ha_{\mu} \, e^{- i \hH t} = \ha_{\mu} \, e^{- i \omega t} $.
Moreover, substituting \eqref{a110} into the Heisenberg equation of motion $\di^2 \hPsi / \di t^2 = \di \hPi / \di t = i \bigl[ \hH, \hPi \bigr]$, we obtain the field equation ${\di^2} \hPsi (\mx,z,t)/{\di t^2} + \omega^2 \hPsi (\mx,z,t) = 0$, which shows that spatial coordinates $\mx=(x,y)$ and time $t$ are uncoupled in a monochromatic field.
Equation \eqref{a110} represents the Hamiltonian of a countably infinite set of identical harmonic oscillators with frequency $\omega$. This type of Hamiltonian is not sporadic in quantum optics; for example, the electromagnetic field within an empty, uniform and nondispersive waveguide with perfectly conducting walls  \cite{doi:10.1119/1.11178}, possesses the very same Hamiltonian.

\begin{figure}[hb!]
  \centering
  \includegraphics[scale=3,clip=false,width=.95\columnwidth,trim = 0 0 0 0]{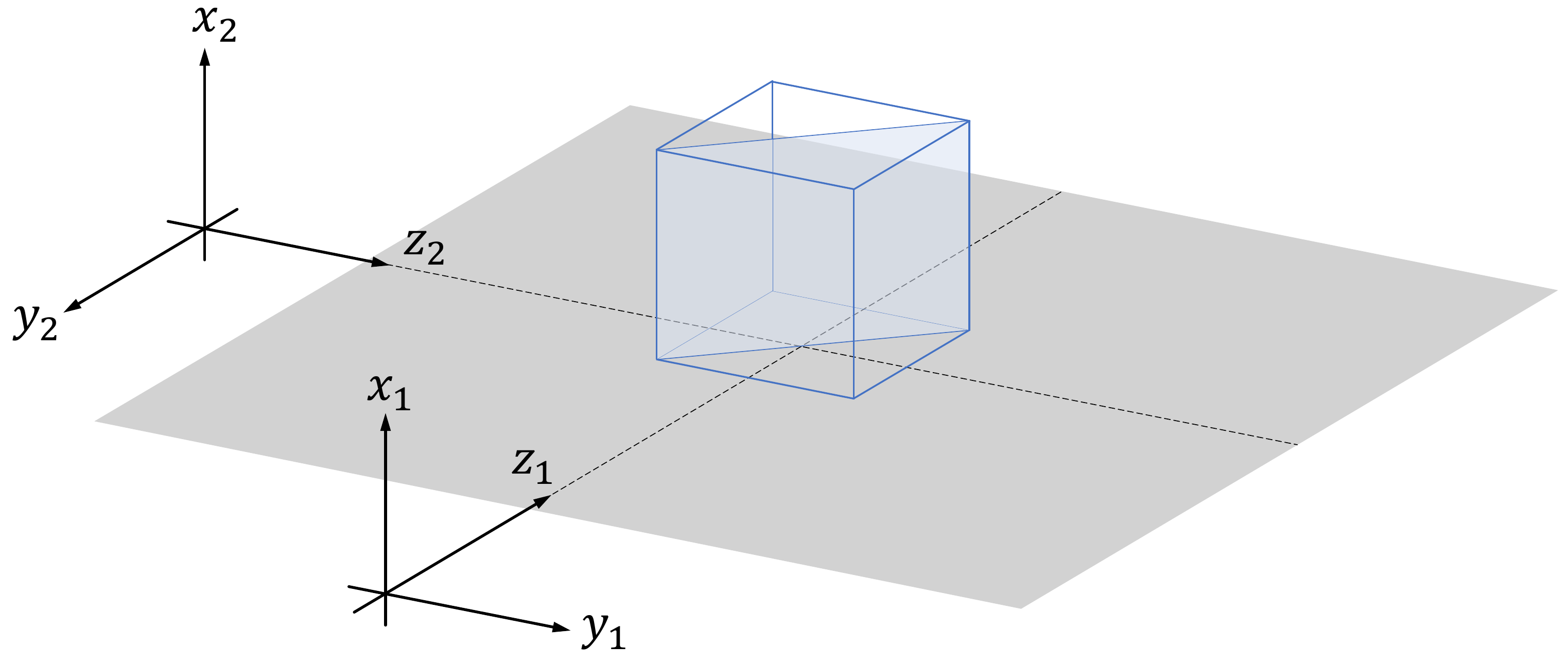}
  \caption{Schematic illustration of a cube beam splitter. This figure shows  the distinct Cartesian coordinate systems $\mr_1 = (\mx_1,z_1)$ and $\mr_2 = (\mx_2,z_2)$  attached to the two input ports of the device. }\label{fig1}
\end{figure}

%%%%%%%%%%%%%%%%%%%%%%%%
\emph{The model.--}
%%%%%%%%%%%%%%%%%%%%%%%%
The geometry of the setup we will consider in the remainder is shown in Fig. \ref{fig1}.
The beam splitter (BS), is characterized by the  reflection and transmission coefficients $\rho$ and $\tau$, respectively, such that \cite{doi:10.1119/1.1432972}:
\begin{align}\label{s260}
\abs{\rho}^2 + \abs{\tau}^2 =1 , \qquad \text{and} \qquad  \rho^* \tau + \rho \, \tau^* = 0.
\end{align}
 The two ports of the BS, here labeled with ``$1$'' and ``$2$'', are aligned with the axes $z_1$ and $z_2$ of two Cartesian coordinates systems $\mr_1=(x_1,y_1,z_1)$ and $\mr_2=(x_2,y_2,z_2)$, respectively. We choose the input electric field entering BS' ports linearly polarized along the (parallel) axes $x_1$ and $x_2$, so that  the scalar description of light \eqref{a70} still applies.
The  quantum fields before and after the BS are described, in the Heisenberg picture, by the Hermitian field operators $\hPsi_\inp$ and $\hPsi_\out$, respectively, defined by (see, e.g., chap. 9 of \cite{GreinerFQ}),
\begin{align}\label{a130}
\hPsi_\as (\mr_1,\mr_2,t) =  \hPsi_{1,\as} (\mr_1,t) + \hPsi_{2,\as} (\mr_2,t),
\end{align}
where ``as'' (asymptotic) refers to either in- or out-fields: ${(\as = \inp, \out)}$, and $\hPsi_\out (\mr_1,\mr_2,t) = \hS^\dagger \hPsi_\inp (\mr_1,\mr_2,t) \hS $, with $\hS$ the unitary scattering operator describing the action of the beam splitter on the input field \cite{Skaar2004}. Note that the single-port fields $\hPsi_{1,\as} (\mr_1,t)$ and $\hPsi_{2,\as} (\mr_2,t)$ are written in the two different Cartesian coordinate systems associated with the two ports of the BS, %with $\mr_i = \mx_i + z_i \, \bm{e}_{z_i}$, $(i=1,2)$.
because we work in the paraxial regime of propagation around two different axes $z_1$ and $z_2$ \cite{PhysRevLett.90.143601}.
We can use (\ref{a70}-\ref{a80}) to write the single-port field $\hPsi_{p,\as} (\mr_p,t)$ as,
\begin{align}\label{a140}
\hPsi_{p,\as} (\mr_p,t) =  \frac{1}{ \sqrt{2 \, \omega}} \left[ \hpp_{p,\as} (\mx_p,z_p,t)  +  \hpm_{p,\as} (\mx_p,z_p,t)  \right],
\end{align}
where
\begin{align}\label{a150}
\hpp_{p,\as} (\mx_p,z_p,t) = e^{- i \omega t} \, \sum_{\mu} \ha_{p\mu, \as} u_{\mu} (\mx_p,z_p)  ,
\end{align}
with $p=1,2$, an index labelling the two ports of the BS.
Using the shorthand $\ha_{p \mu} \equiv \ha_{p\mu, \inp}$ and $\hb_{p\mu} \equiv \ha_{p\mu, \out}$, we can write the BS transformation as \cite{LoudonBook,SupMat},
\begin{subequations}\label{a160}
\begin{align}
\hb_{1\mu} = & \;  \tau \, \ha_{1\mu} + \rho \, (-1)^{m_\mu} \, \ha_{2\mu} , \label{a160A}\\[4pt]
\hb_{2\mu} = & \;   \rho \, (-1)^{m_\mu} \, \ha_{1\mu} + \tau \, \ha_{2\mu}  , \label{a160B}
\end{align}
\end{subequations}
where $\mu = (n_\mu,m_\mu)$, and the factor $(-1)^{m_\mu} $ yields the sign-inversion of the $y$-coordinate of the HG modes due to the reflection at the beam splitter \cite{PhysRevLett.90.143601}.
By construction,
$\bigl[ \ha_{p \mu}, \; \ha^\dagger_{p' \mu'} \bigr] = \delta_{p p'} \delta_{\mu \mu'} = \bigl[ \hb_{p \mu}, \; \hb^\dagger_{p' \mu'} \bigr]$, with $p,p'=1,2$.

Consider now a beam characterized by the field  $\phi(\mx,z) = \varphi(\mx,z)\exp(i k z)$, where $\varphi(\mx,z)$ is a solution of the paraxial wave equation,  normalized according to $\left( \phi, \phi \right) =1$,
where  we have introduced the suggestive notation $(f, g) = \int \di^2 x \, f^*(\mx,z) g(\mx,z) $.
The function $\phi(\mx,z)$ can be  written in terms of the Hermite-Gauss mode functions $u_\mu(\mx,z)$, as
$\phi(\mx,z) = \sum_\mu \phi_\mu u_\mu(\mx,z)$, where $\phi_\mu =  \left( u_\mu, \phi \right)$.
Next, suppose  to have prepared at $t=0$, the field $\phi(\mx,z)$ in the single-photon input  state \cite{Deutsch1991},
\begin{align}\label{a170}
\ket{1[\phi]}^\inp \equiv & \;\ket{1[\phi]}^\inp_1 \ket{0}_2  \nonumber \\[4pt]
= & \;  \biggl( \sum_\mu \phi_{\mu}  \had_{1 \mu} \biggr)\ket{0} \equiv \had_{1}\left[ \phi \right] \ket{0},
\end{align}
which enters the BS from port $1$, while port $2$ is fed with vacuum.
Here and hereafter, the input (output) vacuum state $\ket{0}$ is defined by $\ha_{p \mu} \ket{0} = 0$ ($\hb_{p \mu} \ket{0} = 0$), for all $p$ and $\mu$, and $\ket{0}$ will denote both $\ket{0}^\inp$ and $\ket{0}^\out$.
A straightforward calculation yields
\begin{align}\label{a170bis}
\left[ \ha_{p}\left[ \phi \right] , \had_{p'}\left[ \phi' \right] \right] = \delta_{p p '} \left( \phi, \phi'\right), \qquad (p,p'=1,2).
\end{align}
Note that the field function $\phi(\mx,z)$ determines the so-called  wave function of the input photon, according to
\begin{align}\label{a190}
\mean{0}{\hPsi_{p,\inp} (\mx_p,z_p,t)}{1[\phi]}^\inp = \frac{1}{\sqrt{2 \omega}} \, \phi(\mx_p,z_p) e^{-i \omega t}  .
\end{align}

To characterize the output field after the BS, we need to calculate the eigenstate $\ket{\Psi, t}^\out = \ket{\Psi_1, \Psi_2, t}^\out$ of the Hermitian field operator $\hPsi_\out (\mr_1, \mr_2, t)$ defined by \eqref{a130},
associated with the (time-independent) eigenvalue $\Psi(\mr_1,\mr_2) =  \Psi_1(\mr_1) + \Psi_2(\mr_2)$. From \eqref{a130} it follows that $\ket{\Psi, t}^\out =\ket{\Psi_1, \Psi_2, t}^\out = \ket{\Psi_1, t}^\out_1 \ket{\Psi_2, t}^\out_2$, where
\begin{align}\label{a200}
\hPsi_{p,\out} (\mr_p,t) \ket{\Psi_p, t}^\out_p = \Psi_p(\mr_p)\ket{\Psi_p, t}^\out_p ,
\end{align}
with $p=1,2$ and $\Psi_p(\mr_p)$  a real-valued smooth function, square integrable in the $xy$-plane. It is possible to show \cite{SchwartzBook,SupMat} that the eigenstate $\ket{\Psi_p, t}^\out_p$ is given by
\begin{align}\label{a210}
\ket{\Psi_p, t}^\out_p = & \, \brak{0}{\Psi_p, t}^\out_p \nonumber \\[4pt]
 &  \times  \exp \, \Biggl(   \int \di^2 x  \biggl\{ \sqrt{2 \omega} \, \Psi_p(\mx, z_p) \hpm_{p,\out}(\mx, z_p,t) \nonumber \\[4pt]
 &  \phantom{\times} -\frac{1}{2} \left[\hpm_{p,\out}(\mx, z_p,t)\right]^2 \biggr\} \Biggr) \ket{0},
\end{align}
where the vacuum-field amplitude is given by
\begin{align}\label{a212}
\brak{0}{\Psi_p, t}^\out_p =  \frac{\exp \left[ - \omega \bigl( \Psi_p, \Psi_p \bigr)/2   \right]}{\displaystyle \left\{\int {\mathcal{D} \Psi_p} \, \exp \left[ - \omega \bigl( \Psi_p, \Psi_p \bigr) \right]\right\}^{1/2}}  ,
\end{align}
with $\mathcal{D} \Psi_p$ the functional measure \cite{HatfieldBook,Jackiw1989b}, and $\bra{0}  \hpm_{p,\out}(\mx, z_p,t) =0$, (see Supplemental Material \cite{SupMat} for additional analysis details). The presence of a nonzero electric field in the vacuum state (by definition the right-hand side of \eqref{a212} is always nonzero) is not surprising, being it the analogous of the Gaussian wave function $\varphi_0(q) = \brak{q}{\varphi_0}$ of the ground state $\ket{\varphi_0}$ of a harmonic oscillator in quantum mechanics \cite{CoTannoBook}.

 The quadratic expression $\bigl[\hpm_{p,\out}(\mx, z_p,t)\bigr]^2$ in the exponential in Eq. \eqref{a210}, reveals the singular nature of $\ket{\Psi_p, t}^\out$. These eigenstates can be seen as the quantum-field analogue of the position eigenstates $\ket{q}$ in quantum mechanics: $\hat{Q} \ket{q} = q \ket{q}$.  Thus, they are not normalizable and do not belong to the Hilbert space $\mathcal{H}$ of the light field.
In fact, $\hPsi_\out (\mr_1, \mr_2, t)$ is not  a proper observable because quantum fields are not operators in $\mathcal{H}$, but  rather operator valued distributions \cite{Haag1992}.
%
%\textcolor{blue}{(p. 45 and p. 56)}.
%
To obtain a bona fide Hermitian operator defined on the vectors in  $\mathcal{H}$, one should smear out $\hPsi_\out (\mr_1, \mr_2, t)$ with a smooth real function $f(\mr_1, \mr_2, t)$. However, for the simple analysis that follows we will not need to consider smeared fields. \\
%%%%%%%%%%%%%%%%%%%%%%%%
\emph{Results.--}
%%%%%%%%%%%%%%%%%%%%%%%%
In the remainder, we analyze the single-photon field after the BS using the field-theoretic techniques developed above.
To begin with, we invert the conjugate of Eqs. \eqref{a160}, to write $\had_1[\phi]$ in terms of the output operators
\begin{align}\label{a250}
\hbd_1[\phi] = \sum_\mu \phi_{\mu}  \hbd_{1 \mu}, \quad \text{and} \quad \hbd_2[\widetilde{\phi}] = \sum_\mu \widetilde{\phi}_{\mu}  \hbd_{2 \mu},
\end{align}
where  $\widetilde{\phi}(\mx,z) = \phi(x,-y,z)$. Using this result to convert the input state \eqref{a170} to the corresponding output state, we readily obtain
\begin{align}\label{q40main}
\ket{1[\phi]}^\inp  =  \tau \,  \ket{1[\phi]}_1^\out \ket{0}_2 + \rho \, \ket{0}_1\ket{1[\tilde{\phi}]}_2^\out ,
\end{align}
where $\ket{1[\psi]}_p^\out = \hbd_p[\psi] \ket{0}$, with $\psi = \phi, \widetilde{\phi}$, and the subscript $p=1,2$ label the two output ports of the BS.
Clearly, Eq. \eqref{q40main} gives the \emph{particle representation} of the state $\ket{1[\phi]}^\inp$, because the latter is written in terms of photon number Fock states. However, using the eigenstates \eqref{a210} of the electric field  $\{\ket{\Psi_1,t}^\out_1,  \ket{\Psi_2,t}^\out_2 \}$  as basis vectors, we can write the \emph{field representation}  of the same state $\ket{1[\phi]}^\inp$, as \cite{SupMat},
\begin{align}\label{a260}
\ket{1[\phi]}^\inp   = & \;  \int  \mathcal{D} \Psi_1 \mathcal{D} \Psi_2 \Bigl\{  \/^\out\brak{\Psi_1, \Psi_2,t}{1[\phi]}^\inp \nonumber \\[4pt]
& \phantom{\;  \int  \mathcal{D} \Psi_1 \mathcal{D} \Psi_2 \Bigl\{ } \times  \ket{\Psi_1,t}^\out_1 \ket{\Psi_2,t}^\out_2 \Bigr\}  ,
\end{align}
where
\begin{align}\label{a262}
\/^\out\brak{\Psi_1, \Psi_2,t}{1[\phi]}^\inp = & \;  \sqrt{2 \omega}  \left[\tau \, \bigl( \Psi_1, \phi \bigr) + \rho \, \bigl( \Psi_2, \widetilde{\phi}  \bigr) \right] \nonumber \\[4pt]
& \times  \/^\out\brak{\Psi_1,\Psi_2,t}{0} \, e^{- i \omega t} \, .
\end{align}
The state vector $\ket{1[\phi]}^\inp$ is obviously the same in both Eqs. \eqref{q40main} and \eqref{a260}, but its representation is not. The particle description of $\ket{1[\phi]}^\inp$ forced by \eqref{q40main} is vanished in the field representation given by \eqref{a260} (see also \cite{vedral2021quantum} for a clear discussion on this point).

\begin{figure}[h!]
  \centering
  \includegraphics[scale=3,clip=false,width=.9\columnwidth,trim = 0 0 0 0]{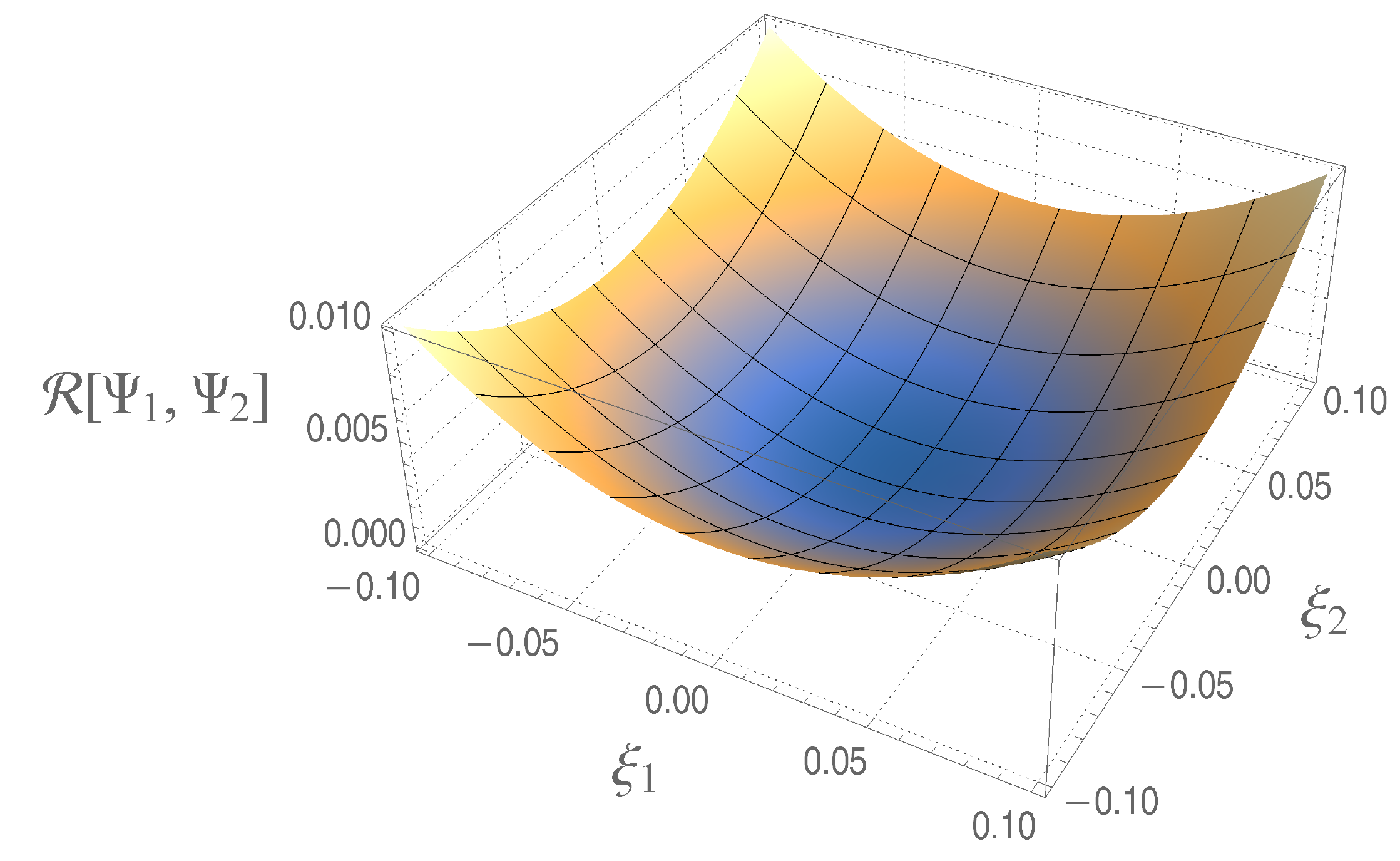}
  \caption{Equation \eqref{a238} is plotted as a function of the (small) dimensionless displacements $\xi_1$ and $\xi_2$. The value $\mathcal{R}[\Psi_1,\Psi_2] =0$ is attained only when both $\xi_1=0$ and $ \xi_2=0$, occurring when both  the measured fields $\Psi_1$ and $\Psi_2$ are orthogonal to the single-photon wave functions $\phi$ and $\widetilde{\phi}$. }\label{fig2}
\end{figure}

Now, given the input state $\ket{1[\phi]}^\inp$ at time $t=0$,  we can ask what is the probability  $\mathcal{D} P(t)$ that the observation of the output electric field at a later time $t>0$ will yield a value centered around $\Psi(\mr_1, \mr_2) =  \Psi_1(\mr_1) + \Psi_2(\mr_2)$ within $\mathcal{D} \Psi_1 \mathcal{D} \Psi_2$.
Formally, such probability is given by $\mathcal{D} P(t) = \abs{\/^\out\brak{\Psi_1, \Psi_2,t}{1[\phi]}^\inp}^2 \mathcal{D} \Psi_1 \mathcal{D} \Psi_2$.
Practically, from \eqref{a262} it follows that we can simultaneously observe a nonzero electric field behind both ports $1$ and $2$ of the BS, and that the value of this field is determined by the input single-photon field $\phi$ via the amplitudes  $\tau \, \sqrt{2\omega}\bigl( \Psi_1, \phi \bigr)$ and $  \rho \, \sqrt{2\omega}\bigl( \Psi_2, \widetilde{\phi}  \bigr)$, respectively. By choosing $\phi \in \mathbb{R}$ and $\Psi_1 = \phi/\sqrt{2\omega}$ and $\Psi_2 = \widetilde{\phi}/\sqrt{2\omega}$, we can maximize the functional $\mathcal{R}[\Psi_1,\Psi_2]$ defined by
\begin{align}\label{a237}
\mathcal{R}[\Psi_1,\Psi_2] \equiv & \;  \abs{\frac{\/^\out\brak{\Psi_1, \Psi_2,t}{1[\phi]}^\inp}{\/^\out\brak{\Psi_1,\Psi_2}{0}}}^2 \nonumber \\[4pt]
= & \; 2 \omega \abs{\tau \, \bigl( \Psi_1, \phi \bigr) + \rho \, \bigl( \Psi_2, \widetilde{\phi}  \bigr)}^2.
\end{align}
This  means that the most probable field configuration to be measured, is the one coinciding with the field of the input photon, as expected. However, it is more important to note that  while the vacuum-field amplitude \eqref{a212} is always nonzero by definition, there are special field configurations that have zero probability to be measured, thus witnessing the passage of the photon through \emph{both} both ports of the BS. I fact, if we choose  $\Psi_1$ and $\Psi_2$ such that $\bigl( \Psi_1, \phi \bigr) = 0 = \bigl( \Psi_2, \widetilde{\phi} \bigr)$, then $\mathcal{R}[\Psi_1,\Psi_2] = 0$.
For example, suppose to set up an experiment in which we prepare the input single-photon beam in the $\text{TEM}_{00}$ mode by choosing   $\phi(\mx,z) = u_0(\mx,z)  = \varphi_{0}(x,z)
\varphi_{0}(y,z)\exp(i kz) $, where \eqref{a10} has been used. This implies $\phi = \widetilde{\phi}$.
Also, we place the detection apparatuses at the output ports $1$ and $2$ of the BS at $z_1 = 0$ and $z_2=0$, respectively, and we arrange them  in such a way to measure the displaced $\text{TEM}_{10}$ mode   $\sqrt{2 \, \omega} \,\Psi_p(\mx,0) = \varphi_{1}(x + x_0 \, \xi_p,0) \varphi_{0}(y,0), \; (p=1,2)$ in both arms \cite{Magnus2004}, where $\xi_1, \xi_2$ are the (supposedly small) dimensionless displacements of the two measured fields.
  A straightforward calculation gives
\begin{align}\label{a238}
\mathcal{R}[\Psi_1,\Psi_2] = \frac{1}{2} \left( \xi_1^2 \, e^{-\xi_1^2/2} + \xi_2^2 \, e^{-\xi_2^2/2}\right),
\end{align}
for a ${50 \! : \! 50}$ BS with $\tau = 1/\sqrt{2}$ and $\rho = i/\sqrt{2}$.
As illustrated by Fig. \ref{fig2}, we obtain $\mathcal{R}[\Psi_1,\Psi_2] = 0$ only when both $\xi_1 = 0$ and $ \xi_2 = 0$, that is, only when both measured fields $\Psi_1$ and $\Psi_2$ are orthogonal to the single-photon wave functions $\phi$ and $\widetilde{\phi}$ at the two ports of the BS.
\\
This analysis  shows clearly that it is the photon field $\phi$  which is felt at two spatially separated detectors. We remark that such key role of the field $\phi$ cannot be explicitly seen using the simple two-mode state \eqref{a5}, for this we had to use quantum field theory. This is our main result.

Another undoubted advantage of the field representation over the particle one becomes apparent when we consider correlation functions.
For example, the photon-number correlation function at the two outputs of the BS does not, in fact, provide any useful information because, evidently, $\/^\inp\mean{1[\phi]}{\hN_{1,\out}[\phi_1]\hN_{2,\out}[\phi_2]}{1[\phi]}^\inp = 0$,
where $\hN_{p,\out}[\phi_p] = \hbd_p[\phi_p] \hb_p[\phi_p]$,  is the photon-number operator at the output port $p = 1,2$ of the BS, and \eqref{q40main} has been used. However, the two-point field correlation function
\begin{multline}\label{a239bis}
 \/^\inp\bra{1[\phi]}  {\hPsi_{1,\out}(\mr_1,t)\hPsi_{2,\out}(\mr_2,t)}\ket{1[\phi]}^\inp   \\[4pt]
  =  \frac{1}{2 \omega} \Bigl[ \tau \, \rho^* \phi(\mr_1)  \widetilde{\phi}^*(\mr_2)   + \tau^*  \rho \, \phi^*(\mr_1)\widetilde{\phi}(\mr_2) \Bigr],
\end{multline}
is proportional to the product of the single-photon wave functions \eqref{a190} evaluated at $\mr_1=(\mx_1,z)$ and $\mr_2=(\mx_2,z)$ at the output ports $1$ and $2$, respectively, and it is generically nonzero. An exception occurs when $\phi \in \mathbb{R}$, so that the right-hand side of \eqref{a239bis} becomes equal to zero because of the rightmost of Eqs. \eqref{s260}.
%

%%%%%%%%%%%%%%%%%%%%%%%%
\emph{Discussion.--}
%%%%%%%%%%%%%%%%%%%%%%%%
We have seen above that quantum mechanics offers us at least two different representations of a single-photon state. The first one is the \emph{particle representation} given by the Fock state \eqref{q40main}, and the second one is the \emph{field representation} in terms of the eigenstates of the electric field \eqref{a260}. Both equations \eqref{q40main}  and \eqref{a260} are perfectly valid descriptions of the single-photon state  and there are not fundamental physical reasons to choose, a priori, one or the other form. The multiplicity of  equivalent representations of the same state of a physical system is not peculiar to quantum mechanics. Any theory in which the superposition principle is valid presents the same characteristic. For example, in classical optics we can write a light field indifferently as a superposition of  either plane  or spherical waves \cite{Zangwill},  depending on the geometry of the problem. However, no one in classical optics talk about ``plane-spherical duality'', or do really believes that there are truly existing plane or spherical waves in the field.
%So why, in quantum mechanics, should we think in a different way and consider waves or particles with which we %represent the state of a system to be real?

When we have calculated the probability $\mathcal{D} P(t)$ we have claimed that given the input state $\ket{1[\phi]}^\inp$ entering a BS, \emph{we could observe} the fields $\Psi_1(\mr_1, t)$ and $\Psi_2( \mr_2,t)$ behind ports $1$ and $2$ of the BS, respectively.  However, this is not the same as saying that \emph{there are} the fields $\Psi_1(\mr_1, t)$ and $\Psi_2( \mr_2,t)$ after the BS \emph{before} the measurement took place. For, if this were the case, the light field after the first beam splitter would be represented by the state $\ket{\Psi_1, \Psi_2,t}^\out$ and we could detect more than one photon in each port.
Indeed, a straightforward calculation shows that the probability (relative to the vacuum) to  detect $N_p$ photons in the light field of amplitude $\phi_p(\mx_p,z_p)$ at output port $p=1,2$, given the state $\ket{\Psi,t}^\out=\ket{\Psi_1, \Psi_2,t}^\out$, is
\begin{multline}\label{a245}
\abs{\frac{\/^\out\brak{N_1[\phi_1],N_2[\phi_2]}{\Psi_1, \Psi_2, t}^\out}{\brak{0}{\Psi_1,\Psi_2, t}^\out}}^2 \\[4pt]
=  \frac{\text{H}_{N_1}^2 \bigl( \sqrt{ \omega}\left( \Psi_1,\phi_1 \right) \bigr)  \text{H}_{N_2}^2 \bigl( \sqrt{ \omega}\left( \Psi_2,\phi_2 \right) \bigr)}{2^{N_1+N_2} N_1!N_2!},
\end{multline}
where $\/^\out\bra{N_1[\phi_1],N_2[\phi_2]} = \prescript{\out}{1}{\bra{N_1[\phi_1]}} \prescript{\out}{2}{\bra{N_2[\phi_2]}}$, with $\prescript{\out}{p}{\bra{N_p[\phi_p]}}  =  \bra{0} \,\bigl(\hb_{p} [\phi_p] \bigr)^{N_p}/\sqrt{N_p!} \,$,  and $\text{H}_{N_p}(x)$ denotes the  $N_p$th-order Hermite polynomial, $(N_p= 0,1,\ldots, \infty)$.
For the sake of clarity, in Eq. \eqref{a245} we have chosen real-valued photon fields $\phi_1$ and $\phi_2$ normalized  such that  $\left( \phi_1^*,\phi_1 \right)=\left( \phi_2^*,\phi_2 \right)=1$.
Clearly, the quantity \eqref{a245} is equal to zero only in the extraordinary case in which the argument $\sqrt{\omega}\left( \Psi_p,\phi_p \right) $ of the Hermite polynomial $\text{H}_{N_p}$ coincides with a root of the latter. Therefore, if the field were in the state $\ket{\Psi,t}^\out$ there would be a nonzero probability to detect more than one photon, thus contradicting the results of the first experiment reported in \cite{Grangier_1986}.
Hence, if \emph{before} the measurement the field after the BS cannot be represented by the state vector $\ket{\Psi,t}^\out$ but it is found in the state $\ket{\Psi,t}^\out$ \emph{after} the measurement, we must conclude that such a state is created \emph{at} the act of measurement. This point is further discussed in \cite{PhysRevLett.74.4571} and  reflects the fact that often the distinction between preparation and measurement, or test, of a quantum state is subjective \cite{doi:10.1119/1.13586}.

%%%%%%%%%%%%%%%%%%%%%%%%%%%%%%%%%
\emph{Summary.--}
%%%%%%%%%%%%%%%%%%%%%%%%%%%%%%%%%
%
We conclude with a brief summary of our results.
The fundamental experimental setup we have studied coincides, and it could not be otherwise, with the original scheme by Tan \emph{et al.}, and  subsequent versions \cite{PhysRevLett.66.252,PhysRevLett.73.2279,PhysRevA.64.042106}, but our analysis is completely different. Unlike previous studies, we used neither inequalities nor specific measurement techniques to find our results. We simply analyzed very carefully the physics of the problem using  methods and techniques of quantum field theory. Thus, we found that it is the electromagnetic field associated with the single-photon state, rather than the photon itself as a particle, that exhibits nonlocal behavior. The theory presented here is simple in the sense that we have considered the most basic nontrivial model for an optical quantum field, namely a monochromatic, paraxial, scalar field. A more general analysis including perfectly arbitrary electromagnetic fields is perfectly possible by closely following our analysis, but it is just technically more complicated.

I would like to thank Claudiu Genes and Giuliano D'Auria for
stimulating conversations. I am grateful to Gerd Leuchs for providing the bibliographic data of reference \cite{leuchs2015physical}. I acknowledge support from the Deutsche Forschungsgemeinschaft Project No. 429529648-
TRR 306 QuCoLiMa (``Quantum Cooperativity of Light and Matter'').

%\bibliography{all_bibliography_2021}

\end{document}

\title{A Simple Field Theoretic Description of Single-Photon Nonlocality \\ -- Supplemental Material --}

\author{Andrea Aiello}
\email{andrea.aiello@mpl.mpg.de} % optional
\affiliation{Max Planck Institute for the Science of Light, Staudtstrasse 2, 91058
Erlangen, Germany}
% Please provide a full mailing address here.

% See the REVTeX documentation for more examples of author and affiliation lists.

\date{\today}

\maketitle

\thispagestyle{empty}

\tableofcontents

\section{Introduction}

In this Supplemental Material we will give all the details of the calculations that have been  omitted in the main text for brevity. However, much further didactic material has been added. In fact, we have written this Supplemental Material having in mind especially the students who will read the main text and who may not possess yet all the analytical tools needed for its comprehension. %For the sake of clarity, in this additional text we  have made explicit the dependence of the fields from the coordinate $z$, which is just a parameter, as opposed to the dynamical variables $x$ and $y$, in our theory.
Throughout  this note we choose the units so that $\hbar = 1$, as in the main text.

\section{Canonical commutation relations}

\renewcommand\theequation{S\arabic{equation}}
\setcounter{equation}{0}

The field operator is defined by
\begin{align}\label{s10}
\hPsi (\mx, z ,t) =  \frac{1}{ \sqrt{2 \, \omega}} \left[ \hpp (\mx, z ,t)  +  \hpm (\mx, z ,t)  \right],
\end{align}
where
\begin{align}\label{s20}
\hpp (\mx, z  ,t) = & \; \sum_{\mu} \ha_{\mu} u_{\mu} (\mx, z ) \, e^{- i \omega t} ,
\end{align}
and the paraxial-mode annihilation and creation operators $\ha_{\mu}$ and $\ha^\dagger_{\mu'}$, respectively, satisfy the bosonic commutation relation by definition:
\begin{align}\label{s40}
\bigl[ \ha_{\mu}, \; \ha_{\mu'} \bigr] = 0 =  \bigl[ \had_{\mu}, \; \had_{\mu'} \bigr], \qquad \text{and} \qquad \bigl[ \ha_{\mu}, \; \ha^\dagger_{\mu'} \bigr] = \delta_{\mu \mu'} .
\end{align}
The Hermite-Gauss mode functions $u_{\mu} (\mx, z )$ are defined by
\begin{align}\label{s42}
u_{\mu} (\mx, z ) = \varphi_\mu(\mx,z) \exp(i k z),
\end{align}
where
\begin{align}\label{s44}
 \varphi_\mu(\mx,z) = \varphi_{n_\mu}(x,z) \varphi_{m_\mu}(y,z),
\end{align}
is a solution of the $2$D paraxial wave equation (see Eq. (5.6-19) in \cite{MandelBook}, and \cite{GoodmanBook}),
\begin{align}\label{s46}
\left( \frac{\partial^2}{\partial x^2} + \frac{\partial^2}{\partial y^2} + 2 i k \frac{\partial}{\partial z}\right) \varphi_\mu(\mx,z) = 0,
\end{align}
with $k>0$ the wavenumber, and
\begin{align}\label{s48}
\varphi_{n}(x,z) =& \; \frac{1}{\pi^{1/4}} \frac{1}{ \sqrt{2^n n!}} \, \frac{1}{\sqrt{x_0}} \, \text{H}_n \! \left( { x}/{x_0} \right) e^{-(x/x_0)^2/2} \, e^{\frac{i}{2}\frac{z}{z_0} (x/x_0)^2} \, e^{ - i \left( n + \frac{1}{2} \right) \arctan \left( {z}/{z_0} \right) },
\end{align}
where $\text{H}_n(x)$ the Hermite polynomial of order $n$, $z_0 = k w_0^2/ 2$ and $x_0 =  w_0[(1+ z^2/z_0^2)/2]^{1/2}$, with $w_0>0$ the minimum beam waist.

From \eqref{s20}-\eqref{s40} it trivially follows that
\begin{align}\label{s50}
\left[ \hpp (\mx, z  ,t), \; \hpp (\mx', z  ,t) \right] = 0 = \left[ \hpm (\mx, z  ,t), \; \hpm (\mx', z  ,t) \right],
\end{align}
and, a bit  less trivially, that
\begin{align}\label{s60}
\left[ \hpp (\mx, z  ,t), \; \hpm (\mx', z  ,t) \right] = & \;  \sum_{\mu, \mu'}u_{\mu} (\mx, z ) u_{\mu'}^* (\mx', z ) \left[ \ha_{\mu}, \; \ha^\dagger_{\mu'} \right] \nonumber \\[6pt]
= & \;  \sum_{\mu}u_{\mu} (\mx, z ) u_{\mu}^* (\mx', z )\nonumber \\[6pt]
= & \; \delta^\du (\mx  - \mx'),
\end{align}
where the last line follows from the completeness relation of the Hermite-Gauss mode functions $u_{\mu} (\mx, z ) = \varphi_\mu(\mx,z) \exp(i k z)$.

The conjugated field $\hPi (\mx, z ,t)$ is defined by,
\begin{align}\label{s70}
\hPi (\mx, z ,t) = & \; \frac{\di \hPsi (\mx, z ,t)}{\di t}\nonumber \\[6pt]
= & \; \frac{1}{i} \sqrt{\frac{\omega}{ 2}} \left[  \hpp (\mx,z,t)   -  \hpm (\mx,z,t)  \right].
\end{align}
By construction, the two fields $\hPsi (\mx, z ,t)$  and $\hPi (\mx, z ,t)$ satisfy the equal-time commutation relation
\begin{align}\label{s80}
\left[\hPsi (\mx, z ,t)  , \, \hPi (\mx', z ,t) \right] = & \; \left[\frac{1}{ \sqrt{2 \, \omega}} \left[ \hpp (\mx, z ,t)  +  \hpm (\mx, z ,t)  \right], \frac{1}{i} \sqrt{\frac{\omega}{ 2}} \left[  \hpp (\mx', z ,t)   -  \hpm (\mx', z ,t)  \right] \right] \nonumber \\[6pt]
= & \; \frac{1}{2 i} \left\{ - \left[ \hpp (\mx, z  ,t), \; \hpm (\mx', z  ,t) \right] +  \left[ \hpm (\mx, z  ,t), \; \hpp (\mx', z  ,t) \right] \right\}\nonumber \\[6pt]
= & \; i \left[ \hpp (\mx, z  ,t), \; \hpm (\mx', z  ,t) \right] \nonumber \\[6pt]
= & \; i \, \delta^\du \left( \mx - \mx' \right),
\end{align}
where \eqref{s60} has been used.

\section{The Hamiltonian of the field}

The phenomenological  time-independent  Hamiltonian $\hH$ used in the main text is defined by
\begin{align}\label{s90}
\hH =  \; \frac{1}{2} \int \di^2x  \left[  \hPi^2 (\mx,z,t) + \omega^2 \hPsi^2 (\mx,z, t) \right] .
\end{align}
Substituting \eqref{s10} and \eqref{s70} into \eqref{s90}, we obtain
\begin{align}\label{s100}
\hH = & \; \frac{1}{2} \int \di^2x  \left( \left\{ \frac{1}{i} \sqrt{\frac{\omega}{ 2}} \left[  \hpp (\mx,z,t)   -  \hpm (\mx,z,t)  \right] \right\}^2  + \omega^2 \left\{  \frac{1}{ \sqrt{2 \, \omega}} \left[ \hpp (\mx, z ,t)  +  \hpm (\mx, z ,t)  \right] \right\}^2  \right) \nonumber \\[6pt]
= & \; \frac{\omega}{4} \int \di^2x \left[ \left( - \cancel{\hpp \hpp} + \hpp \hpm + \hpm \hpp - \bcancel{\hpm \hpm} \right)
 + \left( \cancel{\hpp \hpp} + \hpp \hpm + \hpm \hpp + \bcancel{\hpm \hpm} \right)
\right]
\nonumber \\[6pt]
= & \; \frac{\omega}{2} \int \di^2x \left[ \hpp(\mx,z, t) \hpm(\mx,z, t) + \hpm(\mx,z, t) \hpp(\mx,z, t) \right] \nonumber \\[6pt]
= & \; \frac{\omega}{2} \sum_{\mu,\mu'} \left[ \ha_\mu \had_{\mu'} \underbrace{\int \di^2 x \, u_{\mu} (\mx, z ) u_{\mu'}^* (\mx, z )}_{= \, \delta_{\mu \mu'}}  + \had_\mu \ha_{\mu'} \underbrace{\int \di^2 x \, u_{\mu}^* (\mx, z ) u_{\mu'} (\mx, z )}_{= \, \delta_{\mu' \mu}}  \right] \nonumber \\[6pt]
= & \; \frac{\omega}{2} \sum_{\mu} \left( \ha_\mu \had_{\mu} +  \had_\mu \ha_{\mu} \right)\nonumber \\[6pt]
   = & \; \omega  \sum_{\mu} \left( \had_{\mu} \ha_{\mu} + \frac{1}{2} \right),
\end{align}
where the orthogonality of the Hermite-Gauss modes and \eqref{s40}  have been used.

Now we use the \emph{operator expansion theorem} \cite{MandelBook},
\begin{align}\label{s110}
\exp(x \hA) \hB \exp(-x \hA) = \hB + x \bigl[ \hA, \hB \bigr] + \frac{x^2}{2!} \bigl[ \hA, \bigl[ \hA, \hB \bigr] \bigr] + \ldots \; ,
\end{align}
where $\hA$ and $\hB$ are operators and $x$ is a number, to show  that
the Hamiltonian \eqref{s100} is the correct generator of time translations, that is
\begin{align}\label{s120}
\hat{a}_{\mu}(t) = e^{ i \hH t} \, \ha_{\mu} \, e^{- i \hH t} = \ha_{\mu} \, e^{- i \omega t} .
\end{align}
Formally handling the infinite constant $\sum_\mu  \frac{1}{2}$ as if it were a finite number, we can write
\begin{align}\label{s130}
\exp \bigl( i \hH t \bigr) \, \ha_{\mu}  \exp \bigl( -i \hH t \bigr)  = & \;
\exp \left[  (i \omega t)  \sum_{\nu} \left( \had_{\nu} \ha_{\nu} + \cancel{\frac{1}{2}} \right) \right]  \ha_{\mu} \exp \left[ -(i \omega t)  \sum_{\nu} \left( \had_{\nu} \ha_{\nu} + \cancel{\frac{1}{2}} \right) \right] \nonumber \\[6pt]
= & \; \exp(x \hA) \hB \exp(-x \hA),
\end{align}
where we hade defined
\begin{align}\label{s140}
x = i \omega t, \qquad \hA = \sum_{\nu}  \had_{\nu} \ha_{\nu}, \qquad \text{and} \qquad \hB = \ha_{\mu}.
\end{align}
Next, we use the relation
\begin{align}\label{s150}
\bigl[ \hX \hY, \hZ \bigr] = \hX  \bigl[ \hY, \hZ \bigr] + \bigl[ \hX, \hZ \bigr] \hY,
\end{align}
with
\begin{align}\label{s160}
\hX =  \had_{\nu}, \qquad \hY = \ha_{\nu}, \qquad  \text{and} \qquad \hZ = \ha_{\mu},
\end{align}
to show that
\begin{align}\label{s170}
\Bigl[ \hA, \hB \Bigr]  = & \; \Bigl[ \sum_{\nu}  \had_{\nu} \ha_{\nu}, \, \ha_{\mu} \Bigr] \nonumber \\[6pt]
= & \; \sum_{\nu}  \bigl[ \had_{\nu} \ha_{\nu}, \, \ha_{\mu} \bigr] \nonumber \\[6pt]
= & \; \sum_{\nu} \Biggl\{ \had_{\nu} \underbrace{\bigl[ \ha_{\nu}, \ha_{\mu} \bigr]}_{= \, 0} + \ha_{\nu} \underbrace{\bigl[ \had_{\nu}, \ha_{\mu} \bigr]}_{= \, - \delta_{\nu \mu}} \Biggr\} \nonumber \\
= & \;  - \ha_{\mu}\nonumber \\[6pt]
= & \;  - \hB,
\end{align}
where \eqref{s40} has been used. Equation \eqref{s170} implies that
\begin{align}\label{s180}
\bigl[ \hA, \hB \bigr]  =   - \hB, \qquad \bigl[ \hA, \bigl[ \hA, \hB \bigr]\bigr]  =   \bigl[ \hA, -\hB \bigr] =  \hB , \qquad  \bigl[ \hA,\bigl[ \hA, \bigl[ \hA, \hB \bigr]\bigr]\bigr] = \bigl[ \hA, \hB \bigr] =  -\hB , \qquad \text{etc.}
\end{align}
Substituting \eqref{s180} into \eqref{s110}, we readily obtain
\begin{align}\label{s190}
\exp(x \hA) \hB \exp(-x \hA) = & \; \hB - x \, \hB + \frac{x^2}{2!} \, \hB -   \frac{x^3}{3!} \, \hB  + \ldots \; \nonumber \\[6pt]
= & \; \hB \left(1  - x + \frac{x^2}{2!}  -   \frac{x^3}{3!}   + \ldots \right) \nonumber \\[6pt]
= & \; \hB \, \exp(-x)\nonumber \\[6pt]
= & \; \ha_{\mu} \, e^{- i \omega t},
\end{align}
where the definitions in Eq. \eqref{s140} have been used.

Note that even if we had written the Hamiltonian \eqref{s10} in the (deceptively finite) form
\begin{align}\label{s200}
\hH = \frac{\omega}{2} \sum_{\nu} \left( \ha_\nu \had_{\nu} +  \had_\nu \ha_{\nu} \right), \qquad  \text{so that} \qquad \hA =\frac{1}{2}  \sum_{\nu} \left( \ha_\nu \had_{\nu} +  \had_\nu \ha_{\nu} \right),
\end{align}
we still would have found $\bigl[ \hA, \hB\bigr]  =  -\hB$ and Eq. \eqref{s190} would be still valid.

\section{The beam splitter transformation}

\begin{figure}[ht!]
  \centering
  \includegraphics[scale=3,clip=false,width=.8\columnwidth,trim = 0 0 0 0]{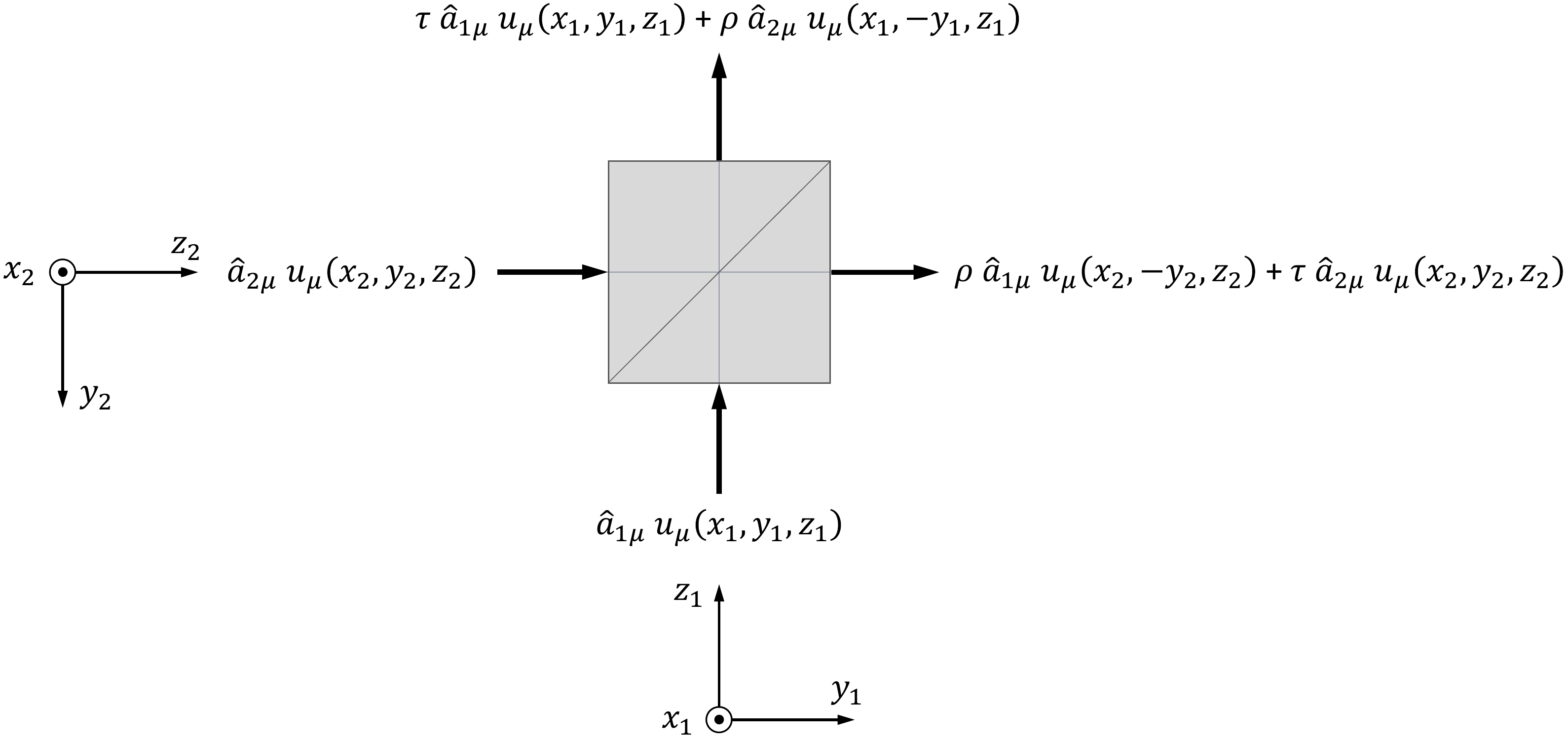}
  \caption{Schematic illustration of a beam splitter. In this figure the Cartesian coordinate systems $\mr_1 = (\mx_1,z_1)$ and $\mr_2 = (\mx_2,z_2)$ are attached to the two input ports of the device. The coordinate axes $x_1$ and $x_2$ are shown directed towards the viewer. Note that under reflection %$\mr_1 \to \mr_2$ and $\mr_2 \to \mr_1$, and
  $u_{\mu} (x_1,y_1, z_1) \to u_{\mu} (x_2,-y_2, z_2)$ and $u_{\mu} (x_2,y_2, z_2) \to u_{\mu} (x_1,-y_1, z_1)$. }\label{fig1s}
\end{figure}

Consider the input and the output positive frequency part of the fields entering and exiting the beam splitter, defined by
\begin{align}\label{s210}
\hpp_\inp (\mx_1,z_1,\mx_2,z_2 ,t) = e^{- i \omega t} \, \sum_{\mu} \Bigl[\ha_{1\mu} u_{\mu} (\mx_1,z_1) + \ha_{2\mu} u_{\mu} (\mx_2,z_2) \Bigr],
\end{align}
and
\begin{align}\label{s220}
\hpp_\out (\mx_1,z_1,\mx_2,z_2 ,t) = e^{- i \omega t} \, \sum_{\mu} \Bigl[\hb_{1\mu} u_{\mu} (\mx_1,z_1) + \hb_{2\mu} u_{\mu} (\mx_2,z_2) \Bigr],
\end{align}
respectively, where the two terms of both $\hpp_\inp (\mx_1,z_1,\mx_2,z_2 ,t)$ and $\hpp_\out (\mx_1,z_1,\mx_2,z_2 ,t)$ are written in two different Cartesian coordinate systems, $\mr_1 = (\mx_1,z_1)$ and $\mr_2 = (\mx_2,z_2)$, because we work in the paraxial regime of propagation around the two orthogonal axes $z_1$ and $z_2$, as shown in Fig. \ref{fig1s}.

The annihilation and creation operators $\ha_{i\mu}$ and $\ha^\dagger_{{i'} \mu'}$, respectively, of the light field entering ports $i=1,2$ and ${i'} =1,2$  of the beam splitter, by definition satisfy the bosonic commutation relations
\begin{align}\label{s40bis}
\bigl[ \ha_{i\mu}, \; \ha_{{i'}\mu'} \bigr] = 0 =  \bigl[ \had_{i\mu}, \; \had_{{i'}\mu'} \bigr], \qquad \text{and} \qquad \bigl[ \ha_{i\mu}, \; \ha^\dagger_{{i'}\mu'} \bigr] = \delta_{i{i'}}\delta_{\mu \mu'} .
\end{align}
The same must be true for the operators $\hb_{i\mu}$ and $\hbd_{{i'}\mu'}$ for the field exiting the beam splitter, that is
\begin{align}\label{s40ter}
\bigl[ \hb_{i\mu}, \; \hb_{{i'}\mu'} \bigr] = 0 =  \bigl[ \hbd_{i\mu}, \; \hbd_{{i'}\mu'} \bigr], \qquad \text{and} \qquad \bigl[ \hb_{i\mu}, \; \hbd_{{i'}\mu'} \bigr] = \delta_{i{i'}}\delta_{\mu \mu'} \;.
\end{align}
We will verify the validity of Eqs. \eqref{s40ter} at the end of this section.

By definition, the input and output fields are connected by the unitary transformation \cite{Skaar2004}:
\begin{align}\label{s230}
\hpp_\out (\mx_1,z_1,\mx_2,z_2 ,t) = \hS^\dagger \hpp_\inp (\mx_1,z_1,\mx_2,z_2 ,t) \hS.
\end{align}
Substituting \eqref{s220} and \eqref{s210} into the left and right-hand sides, respectively, of \eqref{s230}, we obtain
\begin{align}\label{s240}
e^{- i \omega t} \, \sum_{\mu} \Bigl[\hb_{1\mu} u_{\mu} (\mx_1,z_1) + \hb_{2\mu} u_{\mu} (\mx_2,z_2) \Bigr] = e^{- i \omega t} \, \sum_{\mu} \Bigl[\hS^\dagger \ha_{1\mu} \hS \, u_{\mu} (\mx_1,z_1) + \hS^\dagger \ha_{2\mu} \hS \, u_{\mu} (\mx_2,z_2) \Bigr],
\end{align}
which implies that
\begin{align}\label{s250}
\hb_{1\mu} = \hS^\dagger \ha_{1\mu} \hS , \qquad \text{and} \qquad  \hb_{2\mu} = \hS^\dagger \ha_{2\mu} \hS .
\end{align}
Our next goal is to determine $\hS$ to find explicitly the transformation laws in Eq. \eqref{s250}.

To this end, consider a symmetric beam splitter characterized by the  reflection and transmission coefficients $\rho$ and $\tau$, respectively, such that \cite{doi:10.1119/1.1432972}:
\begin{align}\label{s260}
\abs{\rho}^2 + \abs{\tau}^2 =1 , \qquad \text{and} \qquad  \rho^* \tau + \rho \, \tau^* = 0.
\end{align}
From classical optics theory, we know that a paraxial beam of light impinging on a beam splitter is transmitted with amplitude $\tau$ and  reflected with amplitude $\rho$. Moreover, the reflected part undergoes a parity inversion in the horizontal direction, so that $y_1 \to -y_2$ and $y_2 \to -y_1$ (see Fig. \ref{fig1s}). Obviously, reflection also changes the axis of propagation, so that $z_1 \to z_2$ and $z_2 \to z_1$. This means that the mode functions of the light field $u_{\mu} (x_1,y_1,z_1)$ and $u_{\mu} (x_2,y_2,z_2)$, entering port $1$ and port $2$ of the beam splitter, respectively, transform according to
\begin{subequations}\label{s270}
\begin{align}
u_{\mu} (x_1,y_1,z_1) \; \rightarrow \;  & \; \tau \, u_{\mu} (x_1,y_1,z_1) + \rho \, u_{\mu} (x_2,-y_2,z_2), \label{s270A} \\[6pt]
u_{\mu} (x_2,y_2,z_2) \; \rightarrow \; & \; \rho \, u_{\mu} (x_1,-y_1,z_1)  + \tau \, u_{\mu} (x_2,y_2,z_2). \label{s270B}
\end{align}
\end{subequations}
Since the modes of quantum and classical electromagnetic fields satisfy the same classical wave equations, then both quantum and classical fields must transform in the same way under linear transformations. Therefore, we can determine $\hS$ from Eq. \eqref{s230} by imposing that,
\begin{align}\label{s280}
\hpp_\out (\mx_1,z_1,\mx_2,z_2 ,t) = & \; \hS^\dagger \hpp_\inp (\mx_1,z_1,\mx_2,z_2 ,t) \hS \nonumber \\[6pt]
= & \; \Biggl. \hpp_\inp (\mx_1,z_1,\mx_2,z_2 ,t) \Biggr|_{\substack{u_{\mu} (\mx_1,z_1) \; \rightarrow  \; \tau  \, u_{\mu} (x_1,y_1,z_1) \,  +\,  \rho  \, u_{\mu} (x_2,-y_2,z_2) \\[4pt]
u_{\mu} (\mx_2,z_2) \; \rightarrow   \; \rho  \, u_{\mu} (x_1,-y_1,z_1) \,  + \, \tau \, u_{\mu} (x_2,y_2,z_2)}}.
\end{align}
At this point it is useful to remember that the Hermite polynomials satisfy the parity relation
\begin{align}\label{s290}
\text{H}_n(-x) = (-1)^n \text{H}_n(x),
\end{align}
so that the mode function $u_{\mu}(x,y,z) = \varphi_{n_\mu}(x,z) \varphi_{m_\mu}(y,z)\exp(i kz)$ transforms under reflection as
\begin{align}\label{s300}
u_{\mu}(x,-y,z) = & \;  \varphi_{n_\mu}(x,z) \varphi_{m_\mu}(-y,z)\exp(i kz) \nonumber \\[6pt]
=& \;  \varphi_{n_\mu}(x,z) \left[ (-1)^{m_\mu}\varphi_{m_\mu}(y,z) \right] \exp(i kz)\nonumber \\[6pt]
=& \;   (-1)^{m_\mu} u_{\mu}(x,y,z) .
\end{align}
Substituting \eqref{s300} into \eqref{s280}, we obtain
\begin{align}\label{s310}
\hpp_\out (\mx_1,z_1,\mx_2,z_2 ,t) = & \; \Biggl. \hpp_\inp (\mx_1,z_1,\mx_2,z_2 ,t) \Biggr|_{\substack{u_{\mu} (\mx_1,z_1) \; \rightarrow  \; \tau \, u_{\mu} (x_1,y_1,z_1)\,  + \, \rho \, u_{\mu} (x_2,-y_2,z_2) \\[4pt]
u_{\mu} (\mx_2,z_2) \; \rightarrow   \; \rho \, u_{\mu} (x_1,-y_1,z_1)  \, +\,  \tau \, u_{\mu} (x_2,y_2,z_2) }} \nonumber \\[6pt]
= & \;  e^{- i \omega t}  \sum_{\mu} \Bigl\{\ha_{1\mu} \bigl[ \tau  \,  u_{\mu} (x_1,y_1,z_1) + \rho \,   u_{\mu} (x_2,-y_2,z_2) \bigr] + \ha_{2\mu} \bigl[ \rho \,   u_{\mu} (x_1,-y_1,z_1)  + \tau  \,  u_{\mu} (x_2,y_2,z_2) \bigr] \Bigr\} \nonumber \\[6pt]
= & \; e^{- i \omega t}  \sum_{\mu} \Bigl\{\ha_{1\mu} \bigl[ \tau \, u_{\mu} (\mx_1,z_1) + \rho \,  (-1)^{m_\mu} u_{\mu} (\mx_2,z_2) \bigr] + \ha_{2\mu} \bigl[ \rho \,   (-1)^{m_\mu} u_{\mu} (\mx_1,z_1)  + \tau  \,  u_{\mu} (\mx_2,z_2) \bigr] \Bigr\}\nonumber \\[6pt]
= & \; e^{- i \omega t}  \sum_{\mu} \Bigl\{\bigl[ \tau \, \ha_{1\mu} + \rho \,  (-1)^{m_\mu} \ha_{2\mu} \bigr] u_{\mu} (\mx_1,z_1) + \bigl[ \rho \, \ha_{1\mu}  (-1)^{m_\mu}+  \tau \,   \ha_{2\mu} \bigr]u_{\mu} (\mx_2,z_2) \Bigr\}.
\end{align}
Comparing the right-hand side of \eqref{s220} with the right-hand side of \eqref{s310}, we obtain
\begin{align}\label{s320}
\sum_{\mu} \bigl[\hb_{1\mu} u_{\mu} (\mx_1,z_1) + \hb_{2\mu} u_{\mu} (\mx_2,z_2)\bigr]  = \sum_{\mu} \Bigl\{\bigl[ \tau \, \ha_{1\mu} + \rho \,  (-1)^{m_\mu} \ha_{2\mu} \bigr] u_{\mu} (\mx_1,z_1) + \bigl[ \rho \, (-1)^{m_\mu} \ha_{1\mu} +  \tau \, \ha_{2\mu} \bigr]u_{\mu} (\mx_2,z_2) \Bigr\}.
\end{align}
Then, from \eqref{s320} and the orthogonality of the mode functions it follows that
\begin{subequations}\label{s330}
\begin{align}
\hb_{1\mu} = & \;  \tau \, \ha_{1\mu} + \rho \, (-1)^{m_\mu} \, \ha_{2\mu} , \label{s330A}\\[6pt]
\hb_{2\mu} = & \; \rho \, (-1)^{m_\mu} \, \ha_{1\mu} +  \tau \, \ha_{2\mu} . \label{s330B}
\end{align}
\end{subequations}
To convert input to output light field quantum states across the BS \cite{LoudonBook}, it is useful to invert the relations \eqref{s330}, to obtain
\begin{subequations}\label{s330inv}
\begin{align}
\ha_{1\mu} = & \;  \tau^* \, \hb_{1\mu} + \rho^* \, (-1)^{m_\mu} \, \hb_{2\mu} , \label{s330invA}\\[6pt]
\ha_{2\mu} = & \; \rho^* \, (-1)^{m_\mu} \, \hb_{1\mu} +  \tau^* \, \hb_{2\mu} , \label{s330invB}
\end{align}
\end{subequations}
where the unitary character of the transformation, embodied by Eqs. \eqref{s260}, has been used to write
\begin{align}\label{s260bis}
\tau^* = \frac{\tau}{\tau^2 - \rho^2} , \qquad \text{and} \qquad  \rho^* = -\frac{\rho}{\tau^2 - \rho^2}.
\end{align}
For a ${50 \! : \! 50}$ beam splitter we can take
\begin{align}\label{s340}
\rho = \frac{i}{\sqrt{2}}, \qquad \text{and} \qquad \tau = \frac{1}{\sqrt{2}},
\end{align}
so that Eqs. \eqref{s330} reproduce Eqs. (16) in the main text.

We can finally verify the validity of Eqs.   \eqref{s40ter}. That the first relation $\bigl[ \hb_{i \mu}, \; \hb_{{i'} \mu'} \bigr] = 0 =  \bigl[ \hbd_{i \mu}, \; \hbd_{{i'}\mu'} \bigr]$ is satisfied, it trivially follows from $\bigl[ \ha_{i\mu}, \; \ha_{{i'}\mu'} \bigr] = 0 =  \bigl[ \had_{i\mu}, \; \had_{{i'}\mu'} \bigr]$ and Eqs. \eqref{s330}. To verify the second relation we must calculate $\bigl[ \hb_{i\mu}, \hbd_{{i'}\mu'}\bigr]$.
To begin with, using Eqs. \eqref{s330} it is straightforward to calculate
\begin{align}\label{s350}
\Bigl[ \hb_{1\mu}, \hbd_{1\mu'} \Bigr] = & \; \Bigl[\tau \, \ha_{1\mu} + \rho \, (-1)^{m_\mu} \, \ha_{2\mu}, \; \tau^* \, \had_{1\mu'} + \rho^* \, (-1)^{m_{\mu'}} \, \had_{2\mu'}\Bigr] \nonumber \\[6pt]
 = & \; \abs{\tau}^2  \underbrace{\Bigl[\ha_{1\mu} , \; \had_{1\mu'} \Bigr]}_{ =\, \delta_{\mu \mu'}} + \tau \rho^* (-1)^{m_{\mu'}}  \underbrace{\Bigl[\ha_{1\mu} , \; \had_{2\mu'} \Bigr]}_{ =\, 0}
 + \rho \, \tau^*  (-1)^{m_{\mu}} \underbrace{\Bigl[ \ha_{2\mu} , \; \had_{1\mu'} \Bigr]}_{ =\, 0}
+ \abs{\rho}^2  (-1)^{m_{\mu} + m_{\mu'}} \underbrace{\Bigl[ \ha_{2\mu} , \; \had_{2\mu'} \Bigr]}_{ =\, \delta_{\mu \mu'}}   \nonumber \\[6pt]
 = & \; \underbrace{\left( \abs{\tau}^2  + \abs{\rho}^2 \right)}_{ =\, 1, \; \text{from \eqref{s260}}} \delta_{\mu \mu'} \nonumber \\[6pt]
 = & \; \delta_{\mu \mu'} ,
\end{align}
because $(-1)^{m_{\mu} + m_{\mu'}} \delta_{\mu \mu'} = (-1)^{2 m_{\mu}} \delta_{\mu \mu'} = \delta_{\mu \mu'} $. In the same way we can also calculate
\begin{align}\label{s360}
\Bigl[ \hb_{2\mu}, \hbd_{2\mu'} \Bigr] = & \; \Bigl[\rho \, (-1)^{m_\mu} \, \ha_{1\mu} +  \tau \, \ha_{2\mu}, \; \rho^* \, (-1)^{m_\mu} \, \had_{1\mu} +  \tau^* \, \had_{2\mu} \Bigr] \nonumber \\[6pt]
 = & \;  \abs{\rho}^2  (-1)^{m_{\mu} + m_{\mu'}} \underbrace{\Bigl[\ha_{1\mu} , \; \had_{1\mu'} \Bigr]}_{ =\, \delta_{\mu \mu'}} + \abs{\tau}^2  \underbrace{\Bigl[ \ha_{2\mu} , \; \had_{2\mu'} \Bigr]}_{ =\, \delta_{\mu \mu'}}   \nonumber \\[6pt]
 = & \; \left( \abs{\tau}^2  + \abs{\rho}^2 \right) \delta_{\mu \mu'} \nonumber \\[6pt]
 = & \; \delta_{\mu \mu'} .
\end{align}
Next, we calculate
\begin{align}\label{s370}
\Bigl[ \hb_{1\mu}, \hbd_{2\mu'} \Bigr] = & \; \Bigl[\tau \, \ha_{1\mu} + \rho \, (-1)^{m_\mu} \, \ha_{2\mu}, \; \rho^* \, (-1)^{m_{\mu'}} \, \had_{1\mu'} +  \tau^* \, \had_{2\mu'} \Bigr] \nonumber \\[6pt]
 = & \;  \tau r^* (-1)^{m_{\mu'}}  \underbrace{\Bigl[\ha_{1\mu} , \; \had_{1\mu'} \Bigr]}_{ =\, \delta_{\mu \mu'}}
 + \abs{\tau}^2 \underbrace{\Bigl[ \ha_{1\mu} , \; \had_{2\mu'} \Bigr]}_{ =\, 0}
 + \abs{\rho}^2 (-1)^{m_{\mu} + m_{\mu'}} \underbrace{\Bigl[ \ha_{2\mu} , \; \had_{1\mu'} \Bigr]}_{ =\, 0}
 + \rho \, \tau^*  (-1)^{m_{\mu}}  \underbrace{\Bigl[ \ha_{2\mu} , \; \had_{2\mu'} \Bigr]}_{ =\, \delta_{\mu \mu'}}   \nonumber \\[6pt]
 = & \; \underbrace{\left(\rho^* \tau + \rho \, \tau^* \right)}_{ =\, 0, \; \text{from \eqref{s260}}}  (-1)^{m_{\mu}}  \delta_{\mu \mu'} \nonumber \\[6pt]
 = & \; 0 .
\end{align}
Then, from \eqref{s370} it readily follows that
\begin{align}\label{s380}
\Bigl[ \hb_{2\mu}, \hbd_{1\mu'} \Bigr] = \left(\Bigl[  \hb_{1\mu'}, \hbd_{2\mu} \Bigr]\right)^\dagger = 0 .
\end{align}

\section{Quantum states of the light field}

To begin with, we rewrite the Hamiltonian \eqref{s100} in the normal-ordered form
\begin{align}\label{s390}
\nH = & \; \omega  \sum_{\mu}  \had_{\mu} \ha_{\mu}  \nonumber \\[6pt]
= & \; \omega  \sum_{\mu}  \hN_{\mu} \nonumber \\[6pt]
= & \; \omega  \hN,
\end{align}
where we have defined the standard single-mode and the multi-mode number operators $\hN_{\mu}$ and $\hN$, respectively, as
\begin{align}\label{s400}
\hN_{\mu} =   \had_{\mu} \ha_{\mu}, \qquad \text{and} \qquad \hN = \sum_{\mu}  \hN_{\mu}.
\end{align}
So, from now on, every time we will write $\hH$ we will actually mean $\nH\,$.

Note that although each mode function $u_\mu(\mx,z)$ is characterized by the \emph{pair} of indexes $\mu = (n_\mu, m_\mu)$, it is associated with a \emph{single} harmonic oscillator. This implies that we deal with the standard quantum harmonic oscillator algebra (see, e.g., \cite{Merzbacher}). Therefore, the eigenstates $\ket{N_\mu}$ of the number operator $\hN_{\mu}$ are defined in the usual manner by
\begin{align}\label{s410}
\ket{N_\mu} = \frac{ \left(\had_{\mu} \right)^{N_\mu}}{\sqrt{N_\mu!}} \ket{0},
\end{align}
where $\ket{0}$ is the ground state defined by $\ha_\mu \ket{0} = 0$, for all $\mu$. Then, by definition, the multi-mode state
\begin{align}\label{s420}
\ket{\left\{ N \right\}} = \ket{N_1} \otimes \ket{N_{2}}  \otimes \cdots \otimes \ket{N_\infty} = \ket{N_{1}, N_{2} , \ldots, N_\infty},
\end{align}
is an eigenstate of $\hN$ with eigenvalue $N = N_{1} + N_{2}  + \ldots + N_\infty$, that is
\begin{align}\label{s430}
\hN \ket{\left\{ N \right\}}  = & \;   \bigl( \hN_{1} + \hN_{2}  + \ldots + \hN_\infty \bigr) \ket{N_{1}, N_{2} ,  \ldots, N_\infty} \nonumber\\[6pt]
= & \;  \bigl( N_{1} + N_{2}  + \ldots + N_\infty \bigr) \ket{N_{1}, N_{2} ,  \ldots, N_\infty} \nonumber\\[6pt]
= & \;  N \ket{\left\{ N \right\}} .
\end{align}
Of course, this formula makes sense only if the sum $N = \sum_\mu N_\mu$ is finite.

Note that in this section we will work mainly in the Schr\"{o}dinger picture (SP), where operators are time-independent. We can easily switch between operators in  SP and Heisenberg picture (HP),  labeled by the indices S and H, respectively, using
\begin{align}\label{sh20}
\hO_\text{H}(t) = \hU^\dagger(t-t_0) \hO_\text{S}(t_0) \hU(t-t_0),
\end{align}
where
\begin{align}\label{sh30}
 \hU(t-t_0) = \exp \left[ - i \hH \left( t - t_0 \right)\right].
\end{align}
In the remainder, without loss of generality, we will set $t_0=0$. Let $\hA_\tS$ be  a generic Hermitian operator in the Schr\"{o}dinger representation, and suppose that it satisfies the eigenvalue equation
\begin{align}\label{sh40}
\hA_\tS \, \ket{a'}_\tS = a' \ket{a'}_\tS,
\end{align}
where $a'$ is a time-independent real number. Multiplying \eqref{sh40} from left by $\hU^\dagger(t)$ and using the relation $\hU(t)\hU^\dagger(t) = \hId$, where $\hId$ denotes the identity operator, we obtain
\begin{align}\label{sh50}
\left[ \hU^\dagger(t) \, \hA_\tS \, \hU(t) \right]\left[ \hU^\dagger(t) \ket{a'}_\tS \right] = a' \left[ \hU^\dagger(t) \ket{a'}_\tS \right].
\end{align}
From \eqref{sh20} it follows that we can rewrite \eqref{sh50} as
\begin{align}\label{sh60}
\hA_\tH(t) \left[ \hU^\dagger(t) \ket{a'}_\tS \right] = a' \left[ \hU^\dagger(t) \ket{a'}_\tS \right].
\end{align}
What kind of time-dependent state vector is $\hU^\dagger(t) \ket{a'}_\tS$? In the remainder we will deal only with states of the form
\begin{align}\label{sh70}
 \ket{a'}_\tS = \hB_\tS \ket{0},
\end{align}
where $\hB_\tS$ is some given operator written in the SP. Then, in this case
\begin{align}\label{sh80}
\hU^\dagger(t) \ket{a'}_\tS = & \; \underbrace{\left[\hU^\dagger(t)\hB_\tS \hU(t)\right]}_{= \, \hB_\tH(t)} \underbrace{\hU^\dagger(t) \ket{0}}_{= \, \ket{0}} \nonumber \\[6pt]
= & \; \hB_\tH(t) \ket{0},
\end{align}
and we can rewrite \eqref{sh60} as
\begin{align}\label{sh90}
\hA_\tH(t) \left[ \hB_\tH(t) \ket{0}\right] = a' \left[ \hB_\tH(t) \ket{0} \right].
\end{align}
We will use this formula soon.

\subsection{Single-photon states}

Let $\phi(\mx,z) = \varphi(\mx,z)\exp(i k z) $ be the positive-frequency part of some optical  field, where $\varphi(\mx,z)$ is a solution of the paraxial wave equation, square integrable over the $xy$-plane and normalized to $1$, that is
\begin{align}\label{s435}
\left( \frac{\partial^2}{\partial x^2} + \frac{\partial^2}{\partial y^2} + 2 i k \frac{\partial}{\partial z}\right) \varphi(\mx,z) = 0, \qquad \text{and} \qquad \int \di^2 x  \abs{\varphi(\mx,z)}^2 = 1.
\end{align}
Following \cite{LoudonBook} and \cite{Deutsch1991}, we define the time-independent annihilation and creation operators $\ha[\phi]$ and $\had[\phi]$, respectively, associated with the field $\phi(\mx,z)$, as
\begin{subequations}\label{s440}
\begin{align}
\ha[\phi] = & \;  \int \di^2 x \, \phi^*(\mx,z) \hpp(\mx,z,t=0) , \label{s440A}\\[6pt]
\had[\phi] = & \; \int \di^2 x \, \phi(\mx,z) \hpm(\mx,z,t=0) . \label{s440B}
\end{align}
\end{subequations}
Substituting \eqref{s20} into \eqref{s440A}, we obtain
\begin{align}\label{s450}
\ha[\phi] = & \;  \int \di^2 x \, \phi^*(\mx,z) \hpp(\mx,z,0)  \nonumber \\[6pt]
 = & \;  \sum_{\mu} \ha_{\mu} \int \di^2 x \, \phi^*(\mx,z)  u_{\mu} (\mx, z )  \nonumber \\[6pt]
 = & \;  \sum_{\mu} \ha_{\mu} \left[ \int \di^2 x \,   u_{\mu}^* (\mx, z ) \phi(\mx,z) \right]^* \nonumber \\[6pt]
 = & \;  \sum_{\mu} \ha_{\mu} \phi_\mu^*,
\end{align}
where we have defined the $\mu$-component of the function $\phi(\mx,z) = \varphi(\mx,z) \exp(ikz)$ with respect to the basis $u_\mu(\mx,z) = \varphi_\mu(\mx,z) \exp(ikz)$, as
\begin{align}\label{s460}
\phi_\mu =  \int \di^2 x \,   u_{\mu}^* (\mx, z ) \phi(\mx,z) = \int \di^2 x \,   \varphi_{\mu}^* (\mx, z ) \varphi(\mx,z).
\end{align}
That the coefficients $\phi_\mu$ are independent of $z$, follows from the fact that both $\varphi_\mu(\mx,z)$ and $\varphi(\mx,z)$ satisfy the paraxial wave equation, and from the unitary character of the Fresnel propagator \cite{PhysRevE.59.7152}. From \eqref{s440B} and \eqref{s450} it follows that
\begin{align}\label{s470}
\ha^\dagger[\phi] =  \sum_{\mu} \had_{\mu} \phi_\mu.
\end{align}
Using  \eqref{s450} and \eqref{s470}, we can readily calculate the commutator,
\begin{align}\label{s480}
\Bigl[ \ha[\phi] , \had[\phi'] \Bigr] = & \; \sum_{\mu, \mu'}  \phi_\mu^* \phi_{\mu'}' \underbrace{\Bigl[ \ha_\mu , \had_{\mu'} \Bigr]}_{= \, \delta_{\mu \mu'}} \nonumber \\[6pt]
= & \; \sum_{\mu}  \phi_\mu^* \phi_{\mu}'  \nonumber \\[6pt]
= & \; \sum_{\mu}  \left[ \int \di^2 x \, u^*_{\mu}(\mx,z) \phi(\mx, z)\right]^* \left[ \int \di^2 x'\, u^*_{\mu}(\mx',z) \phi'(\mx', z)\right] \nonumber \\[6pt]
= & \;  \int \di^2 x  \int \di^2 x'\,  \phi^*(\mx, z)\phi'(\mx', z) \underbrace{\sum_{\mu}  u_{\mu}(\mx,z) u^*_{\mu}(\mx',z)}_{= \, \delta^\du(\mx - \mx')}  \nonumber \\
= & \;  \int \di^2 x   \,  \phi^*(\mx, z)\phi'(\mx, z).
% \nonumber \\[6pt]
%
%\equiv & \;  \left(  \phi, \phi' \right) =  \left[  \phi^* \, \phi' \right],
%
\end{align}
In the remainder we will use indifferently either $\left(  f, g \right)$ or $\left[  f^* \, g \right]$, to denote the superposition integral
\begin{align}\label{s482}
 \int \di^2 x   \,  f^*(\mx, z) g(\mx, z) =  \left( f, g \right) =  \left[  f^* \, g \right],
\end{align}
where the functional product notation $\left[  f^* \, g \right]$ is borrowed  from Eq. (12.49) in \cite{MandldBook}. Thus, we can rewrite compactly \eqref{s480} as
\begin{align}\label{s480bis}
\Bigl[ \ha[\phi] , \had[\phi'] \Bigr] =   \left(  \phi, \phi' \right) =  \left[  \phi^* \, \phi' \right],
\end{align}
where, by hypothesis,  $\left(  \phi, \phi \right)  = 1 = \left(  \phi', \phi' \right) $.

Using $\had[\phi]$, we can build the photon-number states, denoted by  $\ket{N[\phi]}$, and  defined by
\begin{align}\label{s490}
\ket{N[\phi]} = \frac{ \left(\had[\phi] \right)^{N}}{\sqrt{N!}} \ket{0}.
\end{align}
It is not difficult to show that
\begin{align}\label{s500}
\hN \ket{N[\phi]} = N \ket{N[\phi]},
\end{align}
where $\hN$ is given by \eqref{s400}. Let us do this calculation  in detail. To begin with, we write
\begin{align}\label{s510}
\hN \ket{N[\phi]} = & \; \sum_\mu \had_\mu \ha_\mu \, \frac{\left(\had[\phi] \right)^N}{\sqrt{N!}}\ket{0} \nonumber \\[6pt]
= & \; \frac{1}{\sqrt{N!}} \, \sum_\mu \Bigl[\had_\mu \ha_\mu , \, \left(\had[\phi] \right)^N \Bigr] \ket{0},
\end{align}
which is correct because $ \left(\had[\phi] \right)^N \had_\mu \ha_\mu\ket{0}  =0$.
Next, we use the relation \cite{Lowell1999},
\begin{align}\label{s520}
\bigl[ \hA, \hB_1 \hB_2 \cdots \hB_N \bigr] = \bigl[ \hA, \hB_1 \bigr] \hB_2 \cdots \hB_N + \hB_1 \bigl[ \hA, \hB_2 \bigr] \hB_3 \cdots \hB_N + \cdots +\hB_1 \cdots \hB_{N-1}\bigl[ \hA,  \hB_N \bigr],
\end{align}
to calculate
\begin{align}\label{s530}
\bigl[ \hA, \hB^N \bigr] = \bigl[ \hA, \hB \bigr] \hB^{N-1}  + \hB \bigl[ \hA, \hB \bigr] \hB^{N-2} +  \cdots \hB^{N-1} \bigl[ \hA,  \hB \bigr].
\end{align}
Tacking $\hA = \had_\mu \ha_\mu$ and $\hB = \had[\phi]$, we first calculate
\begin{align}\label{s540}
\bigl[ \hA, \hB \bigr] = & \;  \bigl[ \had_\mu \ha_\mu, \had[\phi] \bigr] \nonumber \\[6pt]
=  & \;  \sum_{\mu'} \phi_{\mu'} \bigl[ \had_\mu \ha_\mu, \had_{\mu'} \bigr] \nonumber \\[6pt]
=  & \;  \sum_{\mu'} \phi_{\mu'} \biggr\{ \had_\mu \underbrace{[ \ha_\mu, \had_{\mu'} \bigr]}_{= \, \delta_{\mu \mu'}} +
 \underbrace{[ \had_\mu, \had_{\mu'} \bigr]}_{= \, 0} \ha_\mu \biggl\}\nonumber \\
=  & \; \phi_\mu \had_\mu,
\end{align}
where \eqref{s150} has been used.
This implies that $\bigl[ \bigl[ \hA, \hB \bigr], \hB \bigr] = 0$ and, therefore, that
\begin{align}\label{s550}
\bigl[\had_\mu \ha_\mu , \, \left(\had[\phi] \right)^N \bigr] = \bigl[ \hA, \hB^N \bigr] = N \, [ \hA, \hB \bigr] \hB^{N-1}  = N \bigl( \phi_\mu \had_\mu \bigr) \bigl(\had[\phi] \bigr)^{N-1}.
\end{align}
Substituting \eqref{s550} into \eqref{s510}, we obtain
\begin{align}\label{s560}
\hN \ket{N[\phi]} = & \;  N \underbrace{ \left( \sum_\mu \phi_\mu \had_\mu \right)}_{ = \, \had[\phi]} \frac{\bigl(\had[\phi] \bigr)^{N-1}}{\sqrt{N!}} \ket{0} \nonumber \\[6pt]
= & \; N \ket{N[\phi]} ,
\end{align}
which correctly reproduces \eqref{s500}.

The number state $\ket{N[\phi]}$ is normalized according to
\begin{align}\label{s560a}
\brak{N[\phi]}{N'[\phi']} = \left( \phi, \phi'\right)^{N}  \delta_{N N'}.
\end{align}
The demonstration is a straightforward calculation, let us do it assuming, without loss of generality, that $N' \geq N$. First, we write
\begin{align}\label{s560b}
\brak{N[\phi]}{N'[\phi']} = & \; \frac{1}{\sqrt{N N'}}\, \mean{0}{\bigl( \ha[\phi] \bigr)^{N} \bigl( \had[\phi'] \bigr)^{N'}}{0} \nonumber \\[6pt]
= & \; \frac{1}{\sqrt{N N'}}\, \mean{0}{\Bigl[\bigl( \ha[\phi] \bigr)^{N},   \bigl( \had[\phi'] \bigr)^{N'} \Bigr]}{0}.
\end{align}
Next, we use \eqref{s530} with $\hA = \bigl(\ha[\phi] \bigr)^{N}$ and  $\hB = \had[\phi']$, to calculate
\begin{align}\label{s560c}
 \mean{0}{\bigl( \ha[\phi] \bigr)^{N}   \bigl( \had[\phi'] \bigr)^{N'} }{0} = & \;
 \mean{0}{\Bigl[\bigl( \ha[\phi] \bigr)^{N},   \bigl( \had[\phi'] \bigr)^{N'} \Bigr]}{0}\nonumber \\[6pt]
 = & \;
 \mean{0}{\Bigl[ \bigl( \ha[\phi] \bigr)^{N},  \had[\phi']   \Bigr] \bigl( \had[\phi'] \bigr)^{N'-1} }{0}  + \underbrace{\bra{0} \had[\phi']}_{= \, 0} \Bigl[ \bigl(\ha[\phi] \bigr)^{N}, \had[\phi'] \Bigr] \bigl( \had[\phi'] \bigr)^{N'-2} \ket{0} \nonumber \\[6pt]
 & +  \cdots + \underbrace{\bra{0}\bigl( \had[\phi'] \bigr)^{N'-1}}_{= \, 0} \Bigl[ \bigl( \ha[\phi] \bigr)^{N},  \had[\phi'] \bigr) \Bigr] \ket{0}\nonumber \\[6pt]
= & \;  \mean{0}{\Bigl[ \bigl( \ha[\phi] \bigr)^{N},  \had[\phi']   \Bigr] \bigl( \had[\phi'] \bigr)^{N'-1} }{0}
\nonumber \\[6pt]
= & \; - \mean{0}{\Bigl[ \had[\phi'] , \bigl( \ha[\phi] \bigr)^{N} \Bigr] \bigl( \had[\phi'] \bigr)^{N'-1} }{0}.
\end{align}
Then, we use again \eqref{s530} but with $\hA = \had[\phi']$ and  $\hB = \ha[\phi]$, to obtain
\begin{align}\label{s560d}
 \mean{0}{\Bigl[ \had[\phi'] , \bigl( \ha[\phi] \bigr)^{N} \Bigr] \bigl( \had[\phi'] \bigr)^{N'-1} }{0} = & \;
 \mean{0}{\underbrace{\Bigl[ \had[\phi'] , \ha[\phi]  \Bigr] }_{= \, - \left( \phi, \phi'\right)}\bigl( \ha[\phi] \bigr)^{N-1} \bigl( \had[\phi'] \bigr)^{N'-1} }{0} \nonumber \\
 & +
  \mean{0}{\ha[\phi]\Bigl[ \had[\phi'] , \ha[\phi]  \Bigr] \bigl( \ha[\phi] \bigr)^{N-2} \bigl( \had[\phi'] \bigr)^{N'-1} }{0}  \nonumber \\[6pt]
 & +  \cdots +    \mean{0}{\bigl( \ha[\phi] \bigr)^{N-1} \Bigl[ \had[\phi'] , \ha[\phi]  \Bigr]  \bigl( \had[\phi'] \bigr)^{N'-1} }{0}  \nonumber \\[6pt]
= & \; - N \left( \phi, \phi'\right) \mean{0}{\bigl( \ha[\phi] \bigr)^{N-1}   \bigl( \had[\phi'] \bigr)^{N'-1} }{0} .
%\nonumber \\[6pt]
%
%= & \; - N \left( \phi, \phi'\right) \mean{0}{\bigl[\bigl( \ha[\phi] \bigr)^{N-1} ,  \bigl( \had[\phi'] \bigr)^{N'-1} %\Bigr] }{0} ,
%
\end{align}
where \eqref{s480bis} has been used. Substituting \eqref{s560d} into \eqref{s560c}, we obtain
\begin{align}\label{s560e}
 \mean{0}{\bigl( \ha[\phi] \bigr)^{N}   \bigl( \had[\phi'] \bigr)^{N'} }{0} =N \left( \phi, \phi'\right) \mean{0}{\bigl( \ha[\phi] \bigr)^{N-1}   \bigl( \had[\phi'] \bigr)^{N'-1} }{0}.
\end{align}
This is a recursive equation of the form
\begin{align}\label{s560f}
f(N,N') = & \; N \left( \phi, \phi'\right) f(N-1,N'-1) \nonumber \\[6pt]
 = & \; N (N-1 )\left( \phi, \phi'\right)^2 f(N-2,N'-2) \nonumber \\
 & \; \phantom{N (N-1 )\left( \phi, \phi'\right)^2} \vdots \nonumber \\
 = & \; N (N-1 ) \cdots (N - k)\left( \phi, \phi'\right)^{k+1} f(N-k-1,N'-k-1) ,
\end{align}
with $f(N,N') =  \mean{0}{\bigl( \ha[\phi] \bigr)^{N}   \bigl( \had[\phi'] \bigr)^{N'} }{0}$. The iteration stops at $N - k -1 =0$ because for $k = N - 1$
\begin{align}\label{s560g}
f(0,N'-N) = & \; \mean{0}{  \bigl( \had[\phi'] \bigr)^{N' - N} }{0} \nonumber \\[6pt]
 = & \; \left\{
          \begin{array}{ll}
            1, & \; \hbox{if} \; N' = N, \\[6pt]
            0, & \; \hbox{if} \; N' > N,
          \end{array}
        \right. \nonumber \\[6pt]
 = & \; \delta_{N N'}.
\end{align}
Substituting \eqref{s560g} into \eqref{s560f}, we obtain $f(N,N')  = N! \left( \phi, \phi'\right)^{N}  \delta_{N N'}$. Using this formula into \eqref{s560b}, we find the sought result \eqref{s560a},
\begin{align}\label{s560h}
\brak{N[\phi]}{N'[\phi']} = \left( \phi, \phi'\right)^{N}  \delta_{N N'}.
\end{align}

\subsection{Coherent states}

The coherent state $\ket{\alpha, [\phi]}$ is defined by
\begin{align}\label{s570}
\ket{\alpha, [\phi]} = \exp \left\{ \alpha \had[\phi] - \alpha^* \ha[\phi] \right\} \ket{0},
\end{align}
where $\alpha$ is a complex number. Substituting Eqs. \eqref{s440} into \eqref{s570}, we obtain
\begin{align}\label{s580}
\ket{\alpha, [\phi]} = & \; \exp \left\{ \alpha  \int \di^2 x \, \phi(\mr) \hpm(\mr,0) - \alpha^* \int \di^2 x \, \phi^*(\mr) \hpp(\mr,0) \right\} \ket{0} \nonumber \\[6pt]
= & \; \exp \left\{ \int \di^2 x \left[ \alpha \, \phi(\mr) \hpm(\mr,0) - \alpha^*   \phi^*(\mr) \hpp(\mr,0) \right] \right\} \ket{0}.
\end{align}
Using the operator relation (see, Eq. (10.11-9) in Ref. \cite{MandelBook})
\begin{align}\label{s590}
\exp \bigl( x \hA\bigr) f \bigl( \hB \bigr) \exp \bigl( -x \hA \bigr) = f \Bigl( \exp \bigl( x \hA )  \hB  \exp \bigl( -x \hA \bigr) \Bigr),
\end{align}
and \eqref{sh80}, it is not difficult to see that
\begin{align}\label{s600}
 \ket{\alpha, [\phi],t} = & \; \hU^\dagger(t) \ket{\alpha, [\phi]} \nonumber \\[6pt]
 = & \; \hU^\dagger(t) e^{ \int \di^2 x \left[ \alpha \, \phi(\mr) \hpm(\mr,0) - \alpha^*   \phi^*(\mr) \hpp(\mr,0) \right] } \hU(t) \underbrace{\hU^\dagger(t) \ket{0}}_{ = \, \ket{0}} \nonumber \\[6pt]
= & \;  e^{ \int \di^2 x \left[ \alpha \, \phi(\mr) \hU^\dagger(t)\hpm(\mr,0)\hU(t) - \alpha^*   \phi^*(\mr) \hU^\dagger(t)\hpp(\mr,0)\hU(t) \right] } \ket{0} \nonumber \\[6pt]
= & \;  e^{ \int \di^2 x \left[ \alpha \, \phi(\mr) \hpm(\mr,t) - \alpha^*   \phi^*(\mr) \hpp(\mr,t) \right] }\ket{0} ,
%,
%
\end{align}
where
\begin{align}\label{s610}
\hU(t) = & \; \exp \left( - i \hH t \right) \nonumber \\[6pt]
= & \;  \exp \left( - i \omega t \, \sum_\mu \had_\mu \ha_\mu \right) ,
\end{align}
and \eqref{s390} has been used. To pass from the third to the fourth line of Eq. \eqref{s600} we have used the relations
\begin{align}\label{s620}
\hU^\dagger(t)\hA(\mr,0)\hU(t)  = \hA(\mr,t), \qquad \text{and} \qquad \hU^\dagger(t)\hA^\dagger(\mr,0)\hU(t)  = \hA^\dagger(\mr,t),
\end{align}
which at this point should be evident. However, for completeness we perform explicitly the calculation for $\hpp(\mr,0)$ (for $\hpm(\mr,0)$ it is basically the same). We use Eqs. \eqref{s20} to rewrite
%
%\begin{align}\label{s630}
%
%\hU^\dagger(t)\hpm(\mr,0)\hU(t) = & \; \exp \left(  i \omega t \, \sum_{\mu'} \had_{\mu'} \ha_{\mu'} \right)  \left[\sum_{\mu} \ha_{\mu} u_{\mu} (\mx, z )\right]  \exp \left( - i \omega t \, \sum_{\mu''} \had_{\mu''} \ha_{\mu''} \right)
%\nonumber \\[6pt]
%
% = & \; \sum_{\mu} u_{\mu} (\mx, z ) \left\{\exp \left(  i \omega t \, \sum_{\mu'} \had_{\mu'} \ha_{\mu'} \right) \ha_{\mu} \exp \left( - i \omega t \, \sum_{\mu''} \had_{\mu''} \ha_{\mu''} \right)\right\}
%\nonumber \\[6pt]
%
%= & \;  \sum_{\mu} \ha_{\mu} u_{\mu} (\mx, z )e^{-i \omega t}
%\nonumber \\[6pt]
%
%= & \;  \hpm(\mr,t) ,
%
%\end{align}
%
%
\begin{align}\label{s630}
\hU^\dagger(t)\hpp(\mr,0)\hU(t) = & \; \exp \left(  i \hH t  \right)  \left[\sum_{\mu} \ha_{\mu} u_{\mu} (\mx, z )\right]  \exp \left(  i \hH t  \right)
\nonumber \\[6pt]
 = & \; \sum_{\mu} u_{\mu} (\mx, z ) \left\{\exp \left(  i \hH t  \right) \ha_{\mu} \exp \left( -i \hH t  \right)\right\}
\nonumber \\[6pt]
= & \;  \sum_{\mu} \ha_{\mu} u_{\mu} (\mx, z )e^{-i \omega t}
\nonumber \\[6pt]
= & \;  \hpp(\mr,t) ,
\end{align}
where \eqref{s120} has been used.

The coherent state $\ket{\alpha, [\phi],t}$ is an eigenstate of the field operator $\hpp(\mr,t)$ with eigenvalue $\alpha \phi(\mr)$. To prove this, first we write
\begin{align}\label{s640}
\hpp(\mx,z,t) \ket{\alpha, [\phi],t} = & \;  \hpp(\mx,z,t) \,  e^{ \int \di^2 x' \left[ \alpha \, \phi(\mx',z) \hpm(\mx',z,t) - \alpha^*   \phi^*(\mx',z) \hpp(\mx',z,t) \right] }\ket{0} \nonumber \\[6pt]
 = & \; \Bigl[ \hpp(\mx,z,t) , \,  e^{ \int \di^2 x' \left[ \alpha \, \phi(\mx',z) \hpm(\mx',z,t) - \alpha^*   \phi^*(\mx',z) \hpp(\mx',z,t) \right] }\Bigr] \ket{0} \nonumber \\[6pt]
\equiv & \; \bigl[\hA , \, e^{\hB} \bigr] \ket{0},
\end{align}
and
\begin{align}\label{s650}
\bigl[\hA , \, \hB \bigr] = & \;  \left[ \hpp(\mx,z,t)  , \,\int \di^2 x' \left\{ \alpha \, \phi(\mx',z) \hpm(\mx',z,t) - \alpha^*   \phi^*(\mx',z) \hpp(\mx',z,t) \right\} \right] \nonumber \\[6pt]
 = & \; \alpha  \int \di^2 x' \, \phi(\mx',z) \underbrace{\Bigl[ \hpp(\mx,z,t) , \,\hpm(\mx',z,t) \Bigr]}_{= \, \delta^\du( \mx - \mx')}  \nonumber \\[6pt]
\equiv & \;  \alpha  \phi(\mx,z).
\end{align}
Since $\alpha  \phi(\mr)$ is a number, this implies that $\bigl[\hB ,  \bigl[\hA , \, \hB \bigr]  \bigr]=0$.
Using the commutation relation \cite{CoTannoBook}
\begin{align}\label{s660}
\bigl[\hA , \, e^{\hB}  \bigr] = \bigl[\hA , \, \hB \bigr]  e^{\hB}, \qquad \text{if} \qquad  \bigl[\hB ,  \bigl[\hA , \, \hB \bigr]  \bigr]=0,
\end{align}
we can rewrite \eqref{s640} as
\begin{align}\label{s670}
\hpp(\mr,t) \ket{\alpha, [\phi],t} = & \;  \bigl[\hA , \, \hB \bigr]  e^{\hB}\ket{0} \nonumber \\[6pt]
 = & \; \alpha \,  \phi(\mr) \, \ket{\alpha, [\phi],t}.
\end{align}
This prove our assertion.

For practical reasons (we will see an application later), it is sometime useful to consider a pseudo eigenvalue equation of the form
\begin{align}\label{s670a}
\hpp(\mr,t_1) \ket{\alpha, [\phi],t_2} =  f(t_1,t_2)\alpha \,  \phi(\mr) \, \ket{\alpha, [\phi],t_2},
\end{align}
where $f(t_1,t_2)$ is a function to be determined, subjected to the constraint  $f(t,t)=1$. To solve Eq. \eqref{s670a} it is enough to rewrite \eqref{s650} at two different times, that is
\begin{align}\label{s650bis}
\bigl[\hA , \, \hB \bigr] = & \;  \left[ \hpp(\mx,z,t_1)  , \,\int \di^2 x' \left\{ \alpha \, \phi(\mx',z) \hpm(\mx',z,t_2) - \alpha^*   \phi^*(\mx',z) \hpp(\mx',z,t_2) \right\} \right] \nonumber \\[6pt]
 = & \; \alpha  \int \di^2 x' \, \phi(\mx',z) \underbrace{\Bigl[ \hpp(\mx,z,t_1) , \,\hpm(\mx',z,t_2) \Bigr]}_{= \, e^{- i \omega(t_1-t_2)}\delta^\du( \mx - \mx')}  \nonumber \\[6pt]
\equiv & \;  e^{- i \omega(t_1-t_2)} \, \alpha  \phi(\mx,z).
\end{align}
From this equation and \eqref{s670a}, it readily follows that
\begin{align}\label{s670b}
 f(t_1,t_2) = e^{- i \omega(t_1-t_2)}.
\end{align}
Note that the simple form of the two-time commutator
\begin{align}\label{s670c}
\Bigl[ \hpp(\mx,z,t_1) , \,\hpm(\mx',z,t_2) \Bigr] = e^{- i \omega(t_1-t_2)} \delta^\du( \mx - \mx'),
\end{align}
comes from the fact that the frequency $\omega$ is the same for all the harmonic oscillators making the field \eqref{s10}.

The coherent state is normalized because
\begin{align}\label{s680}
\brak{\alpha, [\phi]}{\alpha, [\phi]} = & \; \mean{0}{e^{ \alpha^* \ha[\phi] - \alpha \had[\phi] } \, e^{ \alpha \had[\phi] - \alpha^* \ha[\phi] }}{0} \nonumber \\[6pt]
& \; =1.
\end{align}
To see this, using the Campbell-Baker-Hausdorff identity \cite{MandelBook},
\begin{align}\label{s685}
e^{ \hA + \hB } = e^{ \hA} \, e^{ \hB  }  \, e^{-\frac{1}{2}[\hA,\hB]} = e^{ \hB} \, e^{ \hA  }  \, e^{\frac{1}{2}[\hA,\hB]}, \qquad \text{if} \qquad [\hA,[\hA,\hB]] = 0 = [\hB,[\hA,\hB]],
\end{align}
with $\hA = \alpha \had[\phi]$, $\hB = - \alpha^* \ha[\phi] $ and $[\hA,\hB] = \abs{\alpha}^2$ , we find
\begin{align}\label{s690}
e^{ \alpha \had[\phi] - \alpha^* \ha[\phi] } = e^{ -\frac{1}{2} \abs{\alpha}^2} \, e^{ \alpha \had[\phi]  }  \,e^{ -\alpha^* \ha[\phi]  } = e^{ \frac{1}{2} \abs{\alpha}^2} \, e^{ -\alpha^* \ha[\phi]  }  \,e^{ \alpha \had[\phi]  },
\end{align}
so that we can rewrite \eqref{s680} as
\begin{align}\label{s700}
\brak{\alpha, [\phi]}{\alpha, [\phi]} = & \;  e^{ -\abs{\alpha}^2} \underbrace{\bra{0} e^{ -\alpha \had[\phi]  }}_{= \, \bra{0}}  \,e^{ \alpha^* \ha[\phi]  }\;  e^{ \alpha \had[\phi]  }  \, \underbrace{e^{ -\alpha^* \ha[\phi]  } \ket{0}}_{= \, \ket{0}} \nonumber \\
= & \;  e^{ -\abs{\alpha}^2} \underbrace{\bra{0} e^{ \alpha^* \ha[\phi]  }\;  e^{ \alpha \had[\phi]  }  \ket{0} }_{= \, e^{ \abs{\alpha}^2} }\nonumber \\
= &\; 1,
\end{align}
where we have used again the Campbell-Baker-Hausdorff identity, in the form
\begin{align}\label{s710}
e^{\hB} e^{\hA} = e^{[\hB,\hA]} e^{\hA} e^{\hB},
\end{align}
with $\hB = \alpha^* \ha[\phi]$ and $\hA =  \alpha \had[\phi]$.

By definition, the coherent state $\ket{\alpha, [\phi]}$ is also an eigenstate of $\ha[{\phi'}]$ with eigenvalue $\alpha \left( {\phi'}, \phi \right)$:
\begin{align}\label{co10}
\ha[{\phi'}] \ket{\alpha, [\phi]} = \alpha \left( {\phi'}, \phi \right) \ket{\alpha, [\phi]},
\end{align}
where
\begin{align}\label{co20}
 \left( {\phi'}, \phi \right) = \int \di^2 x \, {\phi'}^*(\mx,z) \phi(\mx,z).
\end{align}
This is easy to prove. First, we rewrite \eqref{s440A} as
\begin{align}\label{co30}
\ha[{\phi'}] = & \; \sum_\mu \ha_\mu {\phi'_\mu}^* \nonumber \\[6pt]
= & \; \sum_\mu \ha_\mu \left( {\phi'}, u_\mu \right)
\nonumber \\[6pt]
= & \; \Bigl( {\phi'}, \sum_\mu \ha_\mu u_\mu \Bigr)\nonumber \\[6pt]
= & \; \Bigl( {\phi'}, \hpp \Bigr).
\end{align}
Next, we use \eqref{s670} to calculate  the sought result, that is
\begin{align}\label{co40}
\ha[{\phi'}] \ket{\alpha, [\phi]}  = & \; \Bigl( {\phi'}, \hpp \Bigr) \ket{\alpha, [\phi]} \nonumber \\[6pt]
= & \; \Bigl( {\phi'}, \hpp \ket{\alpha, [\phi]} \Bigr)
\nonumber \\[6pt]
= & \; \Bigl( {\phi'}, \alpha \phi  \ket{\alpha, [\phi]} \Bigr) \nonumber \\[6pt]
= & \; \alpha \bigl( {\phi'}, \phi \bigr)\ket{\alpha, [\phi]} .
\end{align}

We will see later that it is very useful to write the coherent state $\ha[\psi] \ket{\alpha, [\phi]}$ in terms of the number states $\ket{N[\phi]}$ defined by \eqref{s940}. Using the Campbell-Baker-Hausdorff identity \eqref{s690}, we  rewrite \eqref{s570} as
\begin{align}\label{co50}
\ket{\alpha, [\phi]} = & \; e^{ \alpha \had[\phi] - \alpha^* \ha[\phi] } \ket{0} \nonumber \\[6pt]
 = & \; e^{-\abs{\alpha}^2/2}e^{ \alpha \had[\phi]  } \underbrace{e^{ - \alpha^* \ha[\phi] } \ket{0}}_{= \, \ket{0}} \nonumber \\
 = & \; e^{-\abs{\alpha}^2/2} \sum_{N=0}^\infty \frac{\alpha^N}{\sqrt{N!}}\underbrace{\left[ \frac{\bigl(\had [\phi] \bigr)^N}{\sqrt{N!}} \ket{0} \right]}_{= \, \ket{N[\phi]}} \nonumber \\
= & \; e^{-\abs{\alpha}^2/2} \sum_{N=0}^\infty \frac{\alpha^N}{\sqrt{N!}} \ket{N[\phi]}.
\end{align}
The term with $N=0$ in \eqref{co50} gives
\begin{align}\label{co60}
\brak{0}{\alpha, [\phi]} =  e^{-\abs{\alpha}^2/2}.
\end{align}

\subsection{Eigenstates of the electric field}

The Hermitian field operator $\hPsi (\mx, z ,t)$, defined by Eq. \eqref{s10} here reproduced,
\begin{align}\label{s720}
\hPsi (\mx, z ,t) =  \frac{1}{ \sqrt{2 \, \omega}} \left[ \hpp (\mx, z ,t)  +  \hpm (\mx, z ,t)  \right],
\end{align}
satisfies the following eigenvalue equation (in the Heisenberg picture),
\begin{align}\label{s730}
\hPsi (\mx, z,t) \ket{\Psi, t} = & \;\Psi(\mx, z)\ket{\Psi, t} \nonumber \\[6pt]
\equiv & \;   \frac{1}{ \sqrt{2 \, \omega}} \, \psi(\mx, z)\ket{\Psi, t},
\end{align}
where the prefactor $1/\sqrt{2 \, \omega}$ has been inserted for dimensional reasons, and  $\psi(\mx, z)$ is a time-independent real-valued function of $\mx$ and $z$. %We will now switch to the Schr\"{o}dinger representation by fixing the time at $t=0$, so that, for example, $\hPsi_\out (\mx, z) $ will stand $\hPsi_\out (\mx, z,t = 0)$, etc. Note, however,  that $\Psi(\mx, z)$ and $\psi(\mx, z)$ are truly time-independent irrespective of the either Heisenberg or Schr\"{o}dinger pictures.

The first step to prove \eqref{s730} is to notice the similarity between the  commutation relation \eqref{s80},
\begin{align}\label{s740}
\Bigl[\hPsi (\mx, z ,t)  , \, \hPi (\mx', z ,t) \Bigr]  = i \, \delta^\du \left( \mx - \mx' \right),
\end{align}
and the functional commutation relation
\begin{align}\label{s750}
\Bigl[\Psi (\mx, z ,t)  , \, \frac{1}{i} \frac{\delta}{\delta \Psi(\mx', z ,t)} \Bigr] = i \, \delta^\du \left( \mx - \mx' \right).
\end{align}
This analogy advises that in the representation in which $\hPsi (\mx, z ,t)$ is diagonal, the momentum field operator $\hPi (\mx, z ,t)$ has the form $-i \delta/\delta \Psi(\mx, z ,t)$ \cite{SchweberBook,HatfieldBook}. This relation occurs also in ordinary quantum mechanics, where the momentum operator $\hP$ has the representation $-i \di / \di q$ in the coordinate basis $\ket{q}$ where the position operator $\hQ$ is diagonal, that is $\hat{Q} \ket{q} = q \ket{q}$.
Now consider a one-dimensional harmonic oscillator with the Hamiltonian
\begin{align}\label{s760}
\hH = \frac{1}{2m} \hP^2 + \frac{m \omega^2}{2} \hQ^2 = \hbar \omega \left(\had \ha +\frac{1}{2} \right),
\end{align}
where
\begin{align}\label{s770}
\ha = \frac{1}{\sqrt{2}} \left( \frac{\hQ}{q_0} + i \, \frac{\hP }{p_0}\right), \qquad \text{with} \qquad q_0 \equiv \sqrt{\frac{\hbar}{m \omega}}, \qquad \text{and} \qquad p_0 \equiv \frac{\hbar}{q_0}=\sqrt{\frac{1}{ m \omega\hbar}}.
\end{align}
Then, it is not difficult to show that (see, e.g., sec. 6.2 in \cite{FeynmanStat})  the eigenstate $\ket{q}$ of the position operator $\hQ$, can be written as
\begin{align}\label{s780}
\ket{q} = \frac{1}{\pi^{1/4}} \, \frac{1}{\sqrt{q_0}} \, \exp \left[\frac{1}{2} \left(\frac{q}{q_0}\right)^2 \right]
\exp \left[-\frac{1}{2} \left( \had - \sqrt{2} \, \frac{q}{q_0}\right)^2 \right] \ket{0}.
\end{align}
At this point the formal analogy between
\begin{align}\label{s785}
\hQ = \frac{q_0}{\sqrt{2}} \left( \ha + \had \right), \qquad \text{and} \qquad \hPsi (\mx, z ,t) =  \frac{1}{ \sqrt{2 \, \omega}} \left[ \hpp (\mx, z ,t)  +  \hpm (\mx, z ,t)  \right],
\end{align}
suggests that we could try to replace $q/q_0$ and $\had$ in  \eqref{s780}, with $\beta \, \psi(\mx, z)$ and $\hpm (\mx,z,t)$, respectively, so that the sought field eigenstate $\ket{\Psi, t}$ could have tentatively the form
\begin{align}\label{s790}
\ket{\Psi, t} =  \frac{1}{Z^{1/2}[\Psi]} \,
\exp \left\{-\frac{1}{2}  \int \di^2 x \, \Bigl[  \hpm (\mx,z,t)  -  \beta \, \psi(\mx, z) \Bigr]^2  \right\} \ket{0},
\end{align}
where $\beta$ is a real number and  $Z^{1/2}[\Psi]$ is a normalization term, typically a functional of $\Psi$. However, multiplying two or more field operators evaluated at the same spatial point $\mx$  yield divergences, which may be avoided spreading out the factors in the product. Therefore, we try a more general expression like
\begin{align}\label{s800}
\ket{\Psi, t} =  \frac{1}{Z^{1/2}[\Psi]}
\exp \left\{-\frac{1}{2}  \int \di^2 x \, \di^2 x' \, \Bigl[  \hpm (\mx,z,t)  - \beta \, \psi(\mx, z) \Bigr] G(\mx,\mx') \Bigl[  \hpm (\mx',z,t)  -  \beta \,\psi(\mx', z) \Bigr]  \right\} \ket{0},
\end{align}
where the number $\beta$ and the function $G(\mx,\mx') $ will be determined by imposing the validity of \eqref{s730} with $\ket{\Psi, t} $ given by \eqref{s800}. We can take $G(\mx,\mx')$ a symmetric function, that is
\begin{align}\label{s810}
G(\mx,\mx') = G(\mx',\mx).
\end{align}
This can be seen by exchanging the dummy variables $\mx$ and $\mx'$, the order of integration and the first and the last term in the product inside the integral:
\begin{multline}\label{s820}
\exp \left\{-\frac{1}{2}  \int \di^2 x \, \di^2 x' \, \Bigl[  \hpm (\mx,z,t)  - \beta \, \psi(\mx, z) \Bigr] G(\mx,\mx') \Bigl[  \hpm (\mx',z,t)  -  \beta \,\psi(\mx', z) \Bigr]  \right\} \ket{0} \\[6pt]
= \exp \left\{-\frac{1}{2}  \int \di^2 x' \, \di^2 x \, \Bigl[  \hpm (\mx',z,t)  - \beta \, \psi(\mx', z) \Bigr] G(\mx',\mx) \Bigl[  \hpm (\mx,z,t)  -  \beta \,\psi(\mx, z) \Bigr]  \right\} \ket{0} \\[6pt]
= \exp \left\{-\frac{1}{2}  \int \di^2 x \, \di^2 x' \, \Bigl[  \hpm (\mx,z,t)  - \beta \, \psi(\mx, z) \Bigr] G(\mx',\mx) \Bigl[  \hpm (\mx',z,t)  -  \beta \,\psi(\mx', z) \Bigr]  \right\} \ket{0}.
\end{multline}
Now rewrite the left-hand side of \eqref{s730} as
\begin{align}\label{s830}
\hPsi (\mx, z,t) \ket{\Psi, t} = \bigl(\hA + \hB \bigr) e^{\hC}  \ket{0} ,
\end{align}
where
\begin{align}\label{s840}
\hA = \frac{1}{ \sqrt{2 \, \omega}}  \, \hpp (\mx, z ,t) , \qquad \hB = \frac{1}{ \sqrt{2 \, \omega}} \, \hpm (\mx, z ,t)  , \qquad \text{and} \qquad \hC = - \frac{1}{2} \int \di^2 x \, \di^2 x' \, \hc(\mx,z,t) G(\mx, \mx') \hc(\mx',z,t),
\end{align}
with
\begin{align}\label{s850}
\hc(\mx,z,t) = \hpm (\mx,z,t)  - \beta \, \psi(\mx, z).
\end{align}
Note that, by definition,
\begin{align}\label{s860}
\left[ \hB , \, \hc(\mx,z,t) \right]  = 0, \qquad \Rightarrow \qquad \left[ \hB , \, e^{\hC} \right]  = 0,
\end{align}
and that
\begin{align}\label{s870}
\hA  \, e^{\hC}  \ket{0} = \left[ \hA , \, e^{\hC}  \right] \ket{0},
\end{align}
because
\begin{align}\label{s890}
\hA    \ket{0} = \frac{1}{ \sqrt{2 \, \omega}}  \sum_{\mu} \underbrace{\ha_{\mu} \ket{0}}_{= \, 0} u_{\mu} (\mx, z ) \, e^{- i \omega t} =0  .
\end{align}

We would like to use Eq. \eqref{s660}, reproduced below,
\begin{align}\label{s900}
\bigl[\hA , \, e^{\hC}  \bigr] = \bigl[\hA , \, \hC \bigr]  e^{\hC}, \qquad \text{if} \qquad  \bigl[\hC ,  \bigl[\hA , \, \hC \bigr]  \bigr]=0,
\end{align}
to rewrite \eqref{s870} as
\begin{align}\label{s910}
\hA  \, e^{\hC}  \ket{0} = \bigl[\hA , \, e^{\hC}  \bigr] \ket{0} = \bigl[\hA , \, \hC \bigr]  e^{\hC} \ket{0} = \bigl[\hA , \, \hC \bigr] \ket{\Psi, t}.
\end{align}
So, let us calculate
\begin{align}\label{s920}
\bigl[\hA , \, \hC \bigr] = & \; \left[ \frac{1}{ \sqrt{2 \, \omega}}  \, \hpp (\mx, z ,t), \, - \frac{1}{2} \int \di^2 x' \, \di^2 x'' \, \hc(\mx',z,t) G(\mx', \mx'') \hc(\mx'',z,t) \right] \nonumber \\[6pt]
= & \; - \frac{1}{ \sqrt{2 \, \omega}} \, \frac{1}{2} \int \di^2 x' \, \di^2 x'' \, G(\mx', \mx'') \left[  \hpp (\mx, z ,t), \, \hc(\mx',z,t)\hc(\mx'',z,t) \right] .
\end{align}
Now we use
\begin{align}\label{s930}
\bigl[ \hX,  \hY \hZ \bigr] =  \bigl[ \hX, \hY \bigr]\hZ + \hY \bigl[ \hX, \hZ \bigr] ,
\end{align}
with
\begin{align}\label{s940}
\hX =  \hpp (\mx, z ,t), \qquad \hY = \hc(\mx',z,t), \qquad  \text{and} \qquad \hZ = \hc(\mx'',z,t),
\end{align}
to obtain
\begin{align}\label{s950}
\left[  \hpp (\mx, z ,t), \, \hc(\mx',z,t)\hc(\mx'',z,t) \right] = \left[  \hpp (\mx, z ,t), \, \hc(\mx',z,t) \right]\hc(\mx'',z,t) + \left[  \hpp (\mx, z ,t), \,\hc(\mx'',z,t) \right]\hc(\mx',z,t).
\end{align}
Next, we must calculate
\begin{align}\label{s960}
\left[  \hpp (\mx, z ,t), \, \hc(\mx',z,t) \right] = & \; \left[  \hpp (\mx, z ,t), \, \hpm (\mx',z,t)  - \beta \, \psi(\mx', z) \right] \nonumber \\[6pt]
= & \; \delta^\du \left( \mx - \mx' \right),
\end{align}
where \eqref{s60} has been used. This implies that
\begin{align}\label{s970}
\left[  \hpp (\mx, z ,t), \, \hc(\mx',z,t)\hc(\mx'',z,t) \right] = \delta^\du \left( \mx - \mx' \right) \hc(\mx'',z,t) + \delta^\du \left( \mx - \mx'' \right) \hc(\mx',z,t).
\end{align}
Substituting \eqref{s970} into \eqref{s920}, we obtain
\begin{align}\label{s980}
\bigl[\hA , \, \hC \bigr] = & \; - \frac{1}{ \sqrt{2 \, \omega}} \, \frac{1}{2} \int \di^2 x' \, \di^2 x'' \, G(\mx', \mx'') \left[  \delta^\du \left( \mx - \mx' \right) \hc(\mx'',z,t) + \delta^\du \left( \mx - \mx'' \right) \hc(\mx',z,t) \right]\nonumber \\[6pt]
= & \; - \frac{1}{ \sqrt{2 \, \omega}} \, \frac{1}{2} \left\{ \int  \di^2 x'' \, G(\mx, \mx'') \hc(\mx'',z,t) +
\int \di^2 x' \,  G(\mx', \mx) \hc(\mx',z,t) \right\}\nonumber \\[6pt]
= & \; - \frac{1}{ \sqrt{2 \, \omega}} \, \int \di^2 x' \,  G(\mx, \mx') \hc(\mx',z,t) ,
%.
%
\end{align}
where the symmetry \eqref{s810} of $G(\mx, \mx')$ has been exploited. From \eqref{s840} and \eqref{s980}  it follows that
\begin{align}\label{s990}
\bigl[\hC ,  \bigl[\hA , \, \hC \bigr]  \bigr]=0,
\end{align}
and, therefore, Eq. \eqref{s910} holds true, that is
\begin{align}\label{s1000}
 \bigl[\hA , \, \hC \bigr] \ket{\Psi, t} = & \; \left[- \frac{1}{ \sqrt{2 \, \omega}}  \int \di^2 x' \,  G(\mx, \mx') \hc(\mx',z,t) \right]\ket{\Psi, t} \nonumber \\[6pt]
= & \;   \left\{- \frac{1}{ \sqrt{2 \, \omega}}  \int \di^2 x' \,  G(\mx, \mx') \left[ \hpm (\mx',z,t)  - \beta \, \psi(\mx', z) \right] \right\}\ket{\Psi, t}.
\end{align}

The last step is to rewrite \eqref{s830} using \eqref{s1000},
\begin{align}\label{s1100}
\hPsi (\mx, z,t) \ket{\Psi, t} = & \; \hA e^{\hC} \ket{0} + \hB  e^{\hC}  \ket{0} \nonumber \\[6pt]
= & \; \left\{- \frac{1}{ \sqrt{2 \, \omega}} \, \int \di^2 x' \,  G(\mx, \mx') \left[ \hpm (\mx',z,t)  - \beta \, \psi(\mx', z) \right] \right\}\ket{\Psi, t} + \frac{1}{ \sqrt{2 \, \omega}} \, \hpm (\mx, z ,t) \ket{\Psi, t}\nonumber \\[6pt]
= & \;  \frac{1}{ \sqrt{2 \, \omega}}\left\{  \int \di^2 x' \,  G(\mx, \mx') \left[ - \hpm (\mx',z,t)  + \beta \, \psi(\mx', z) \right] +  \delta^\du \left( \mx - \mx' \right) \hpm (\mx', z ,t) \right\}\ket{\Psi, t} .
\end{align}
This expression must be set equal to the right-hand side of  \eqref{s730}, that is
%
%\begin{align}\label{s11100}
%
% \frac{1}{ \sqrt{2 \, \omega}}\left\{  \int \di^2 x' \,  G(\mx, \mx') \left[ - \hpm (\mx',z,t)  + \beta \, \psi(\mx', z) \right] +  \delta^\du \left( \mx - \mx' \right) \hpm (\mx', z ,t) \right\}\ket{\Psi, t} =    \frac{1}{ \sqrt{2 \, \omega}} \, \psi(\mx, z)\ket{\Psi, t}.
%
%\end{align}
%
%
\begin{align}\label{s1110}
 \frac{1}{ \sqrt{2 \, \omega}}\left\{  \int \di^2 x' \,   \left[  \delta^\du \left( \mx - \mx' \right)- G(\mx, \mx')  \right]\hpm (\mx', z ,t) + \beta \, G(\mx, \mx') \, \psi(\mx', z)   \right\}\ket{\Psi, t} =    \frac{1}{ \sqrt{2 \, \omega}} \, \psi(\mx, z)\ket{\Psi, t}.
\end{align}
Clearly, this equation reduces to an identity if and only if
\begin{align}\label{s1120}
 G(\mx, \mx') =  \delta^\du \left( \mx - \mx' \right), \qquad \text{and} \qquad \beta = 1.
\end{align}
Substituting \eqref{s1120} into \eqref{s800}, we obtain
\begin{align}\label{s1130}
\ket{\Psi, t} = &\; \frac{1}{Z^{1/2}[\Psi]}
\exp \left\{-\frac{1}{2}  \int \di^2 x \, \Bigl[  \hpm (\mx,z,t)  -  \psi(\mx, z) \Bigr]^2  \right\} \ket{0} \nonumber \\[6pt]
= &\; \frac{1}{Z^{1/2}[\Psi]}
\exp \left\{-\frac{1}{2}  \int \di^2 x \, \Bigl[  \hpm (\mx,z,t)  -  \sqrt{2 \omega} \, \Psi(\mx, z) \Bigr]^2  \right\} \ket{0},
\end{align}
where \eqref{s730} has been used.
So, our attempt to remove the divergence due to the product of two fields at the same spatial point $\mx$ failed, and we must deal with a non-normalizable state vector. For time being we leave $Z^{1/2}[\Psi] = 1$ undetermined, but we will return to the question of the normalization soon.

\subsection{Eigenstates of the conjugate operator}

By definition, the eigenstates $\ket{\Pi,t}$ of the conjugate operator $\hPi(\mx,z,t)$, which is defined by \eqref{s70}, satisfy the eigenvalue equation
\begin{align}\label{s730bis}
\hPi (\mx, z,t) \ket{\Pi, t} =  & \;  \Pi(\mx, z)\ket{\Pi, t} \nonumber \\[6pt]
=  & \;  \sqrt{\frac{ \omega }{2}} \, \pi(\mx, z)\ket{\Pi, t}.
\end{align}
A straightforward calculation reproducing the  steps  from \eqref{s800} to \eqref{s1130}, shows that
\begin{align}\label{s1130bis}
\ket{\Pi, t} = & \;\frac{1}{Z^{1/2}[\Pi]}
\exp \left\{\frac{1}{2}  \int \di^2 x \, \Bigl[  \hpm (\mx,z,t)  + i \,  \pi(\mx, z) \Bigr]^2  \right\} \ket{0} \nonumber \\[6pt]
= &\; \frac{1}{Z^{1/2}[\Pi]}
\exp \left\{\frac{1}{2}  \int \di^2 x \, \Bigl[  \hpm (\mx,z,t)  + i \, \sqrt{\frac{2}{\omega}} \, \Pi(\mx, z) \Bigr]^2  \right\} \ket{0},
\end{align}
where the normalization functional $Z^{1/2}[\Pi]$ is  to be determined.

\subsection{Superposition of states}

In this section we will calculate the following probability amplitudes:
\begin{enumerate}
  \item Superposition with vacuum: $\brak{\Psi, t}{0}$.
  \item Superposition with the coherent state: $\brak{\Psi, t}{\alpha,[\phi]}$.
  \item Superposition with the number state: $\brak{\Psi, t}{N[\phi]}$,
\end{enumerate}
where
\begin{align}\label{s1135}
\bra{\Psi, t} = \frac{1}{Z^{1/2}[\Psi]}
\bra{0} \exp \left\{-\frac{1}{2}  \int \di^2 x \, \Bigl[  \hpp (\mx,z,t)  -  \psi(\mx, z) \Bigr]^2  \right\} .
\end{align}

\subsubsection{Case $1$}

Note that since $ \hpp (\mx,z,t)\ket{0} = 0$, then
\begin{align}\label{s1140}
\brak{\Psi, t}{0} =  & \;
\frac{1}{Z^{1/2}[\Psi]} \, \exp \left\{-\frac{1}{2}  \int \di^2 x \,  \psi^2(\mx, z)   \right\} \nonumber \\[6pt]
=  & \; \frac{e^{-\frac{1}{2}  \left( \psi, \psi \right) }}{Z^{1/2}[\Psi]} = \frac{e^{-\frac{1}{2}  [ \psi^2 ] }}{Z^{1/2}[\Psi]},
%\exp \left\{-\frac{1}{2} \left( \psi, \psi \right)     \right\} , e^{-\frac{1}{2}  \left( \psi, \psi \right) }
%
\end{align}
where the notation introduced in \eqref{s480} has been used, with $\psi(\mx, z) \in \mathds{R}$.

\subsubsection{Case $2$}

In this case Eq. \eqref{s670a} gives
\begin{align}\label{g10}
\hpp(\mr,t) \ket{\alpha, [\phi]} =   \alpha \, e^{-i \omega t}\phi(\mr)  \ket{\alpha, [\phi]}.
\end{align}
Therefore,
\begin{align}\label{g20}
\brak{\Psi, t}{\alpha,[\phi]} = & \; \frac{1}{Z^{1/2}[\Psi]} \, \bra{0} \exp \left\{-\frac{1}{2}  \int \di^2 x \, \Bigl[  \hpp (\mr,t)  -  \psi(\mr) \Bigr]^2  \right\} \ket{\alpha, [\phi]} \nonumber \\[6pt]
= & \; \frac{1}{Z^{1/2}[\Psi]} \,\exp \left\{-\frac{1}{2}  \int \di^2 x \, \Bigl[    \alpha \, e^{-i \omega t}\phi(\mr)   -  \psi(\mr) \Bigr]^2  \right\} \brak{0} {\alpha, [\phi]}\nonumber \\[6pt]
= & \; \frac{1}{Z^{1/2}[\Psi]} \, \exp \left\{-\frac{1}{2}  \int \di^2 x \, \Bigl[    \alpha \, e^{-i \omega t}\phi(\mr)   -  \psi(\mr) \Bigr]^2  \right\} \exp\left(-\frac{1}{2} \abs{\alpha}^2 \right)\nonumber \\[6pt]
= & \; \underbrace{\frac{ e^{-[ \psi^2]/2} }{Z^{1/2}[\Psi]}}_{= \, \brak{\Psi,t}{0}} e^{-\abs{\alpha}^2/2}\left\{ \exp \left[ \alpha \, e^{-i \omega t}\left[ \psi \,\phi \right] -\frac{1}{2} \alpha^2 \, e^{-2i \omega t}  \left[ \phi^2 \right] \right] \right\}\nonumber \\[6pt]
= & \; \brak{\Psi,t}{0} \,  e^{-\abs{\alpha}^2/2}\left\{ \exp \left[ \alpha \, e^{-i \omega t}\left[ \psi \,\phi \right] -\frac{1}{2} \alpha^2 \, e^{-2i \omega t}  \left[ \phi^2 \right] \right] \right\},
\end{align}
where  \eqref{co60} have been used. As always, we are supposing that $\varphi(\mr)=\phi(\mr)\exp(-i k z)$ is a solution of the paraxial wave equation. Note that when the amplitude $\alpha$ of the coherent state goes to $0$, Eq. \eqref{g20} reproduces correctly Eq. \eqref{s1140}.

\subsubsection{Case $3$}

To evaluate $\brak{\Psi, t}{N[\phi]}$, we use the number state representation of the coherent state, given by \eqref{co50}, here reproduced,
\begin{align}\label{co50bis}
\ket{\alpha, [\phi]} = e^{-\abs{\alpha}^2/2} \sum_{N'=0}^\infty \frac{\alpha^{N'}}{\sqrt{N'!}} \ket{N'[\phi]}.
\end{align}
Then, we can calculate again $\brak{\Psi, t}{\alpha,[\phi]}$ as
\begin{align}\label{g30}
\brak{\Psi, t}{\alpha,[\phi]}\,  e^{\abs{\alpha}^2/2} = \sum_{N'=0}^\infty \frac{\alpha^{N'}}{\sqrt{N'!}} \brak{\Psi, t}{N'[\phi]}.
\end{align}
Tacking the derivative with respect to $\alpha$ of both sides of \eqref{g30}, and then putting $\alpha =0$, we readily obtain the sought amplitudes:
\begin{align}\label{g40}
\brak{\Psi, t}{N[\phi]} = \frac{1}{Z^{1/2}[\Psi]} \,\frac{1}{\sqrt{N!}} \frac{\partial^N}{ \partial \, \alpha^N} \left. \left\{\brak{\Psi, t}{\alpha,[\phi]}\,  e^{\abs{\alpha}^2/2}\right\} \right|_{\alpha \, = \, 0} .
\end{align}
Substituting \eqref{g20} into \eqref{g40}, we obtain
\begin{align}\label{g50}
\brak{\Psi, t}{N[\phi]} = & \; \frac{1}{Z^{1/2}[\Psi]} \, \frac{1}{\sqrt{N!}} \frac{\partial^N}{ \partial \, \alpha^N} \left. \exp \left\{-\frac{1}{2}  \int \di^2 x \, \Bigl[   \alpha \,e^{-i \omega t} \phi(\mr)  -  \psi(\mr) \Bigr]^2  \right\} \right|_{\alpha \, = \, 0}\nonumber \\[6pt]
= & \; \underbrace{\frac{ e^{-[ \psi^2]/2} }{Z^{1/2}[\Psi]}}_{= \, \brak{\Psi,t}{0}} \, \frac{1}{\sqrt{N!}} \frac{\partial^N}{ \partial \, \alpha^N} \left. \left\{ \exp \left[ \alpha \, e^{-i \omega t}\left[ \psi \,\phi\right] -\frac{1}{2} \alpha^2 e^{-2i \omega t}\left[ \phi^2 \right]  \right]\right\}\right|_{\alpha \, = \, 0},
\end{align}
where \eqref{g20} has been used. Now notice that the exponential generating function for the Hermite polynomials \cite{HermitePoly},
\begin{align}\label{g60}
\exp \left( 2 x t - t^2 \right) = \sum_{n=0}^\infty \text{H}_n(x) \frac{t^n}{n!},
\end{align}
permits us to rewrite
\begin{align}\label{g70}
\exp \left\{ \alpha \left[ \psi \,\phi \right] e^{-i \omega t} -\frac{1}{2} \alpha^2 \left[ \phi^2 \right] e^{-2i \omega t}  \right\}= \sum_{n=0}^\infty \text{H}_n \left( \frac{\left[ \psi \,\phi \right]}{\sqrt{2 \left[ \phi^2 \right] }} \right) e^{-  i  n \omega t} \left( \frac{\left[ \phi^2 \right]  }{2} \right)^{n/2} \frac{\alpha^n}{n!}.
\end{align}
Here and hereafter the square root of the (possibly) complex number $\left[ \phi^2 \right] $, must be understood as the principal branch of $\sqrt{\left[ \phi^2 \right] }$, that is the positive square root.
Substituting \eqref{g70} into \eqref{g50}, we obtain
%
%%%%%%%%%%%%%%%%%%%%%%%%%%%%%%%%%%%%%%%%%%%%%%%%%%%%
%
%These calculations are verified in the Mathematica file "calculations_HG_1.nb"
%
%%%%%%%%%%%%%%%%%%%%%%%%%%%%%%%%%%%%%%%%%%%%%%%%%%%
%
%
\begin{align}\label{g80}
\brak{\Psi, t}{N[\phi]}  =  \brak{\Psi,t}{0} \, \frac{e^{- i N  \omega t}}{2^N \sqrt{N!}} \, \text{H}_N \left( \frac{\left[ \psi \,\phi \right]}{\sqrt{2 \left[ \phi^2 \right] }} \right)  \left( \sqrt{ 2 \left[ \phi^2 \right]}  \right)^{N} ,
\end{align}
the first few values of which are given in Table \ref{table1}.
\begin{table}[ht!]
\centering
\begin{tabular}{c!{\vrule width 1.2pt}c}
   $\vphantom{\Biggl[}$ $N$          &  \;  $\brak{\Psi, t}{N[\phi]} /\brak{\Psi, t}{0}$  \; \\[2pt]
\noalign{\hrule height 1.2pt}
$\vphantom{\Biggl[}$  $\phantom{xxx} 0 \phantom{xxx}$  & $1$   \\[10pt]
\hline
$\vphantom{\Biggl[}$ $1$ \qquad             & $\displaystyle \left[ \psi \,\phi \right] e^{- i \omega t}$   \\[10pt]
\hline
$\vphantom{\Biggl[}$ $2$ \qquad            & $\displaystyle  \frac{1}{\sqrt{2}} \left( \left[ \psi \,\phi \right]^2 - \left[ \phi^2 \right]\right) e^{- 2 i \omega t} $   \\[10pt]
\hline
$\vphantom{\Biggl[}$ $3$ \qquad      & $\displaystyle  \frac{\left[ \psi \,\phi \right]}{\sqrt{6}} \left( \left[ \psi \,\phi \right]^2 - 3 \left[ \phi^2 \right]\right)e^{- 3 i \omega t} $   \\[10pt]
\noalign{\hrule height 1.2pt}
\end{tabular}
\caption{Expressions of the amplitudes $ \brak{\Psi, t}{N[\phi]} /\brak{\Psi, t}{0}$ calculated from  \eqref{g50} for $0 \leq N \leq 3$.}
\label{table1}
\end{table}
Note that if we choose $\psi$ and $\phi$ such that $\left[\psi \, \phi \right] = \left( \psi , \phi \right)=0$, then $\brak{\Psi, t}{N[\phi]} =0$ for $N$ odd, as for squeezed states.

\section{Input and Output states of the light field}

In the main text we consider three different quantum states of light entering the beam splitter from port $1$ (port $2$ is \emph{always} fed with vacuum): the vacuum state $\ket{0}=\ket{0}_1\ket{0}_2$, the single-photon state $\ket{1[\phi]}= \ket{1[\phi]}_1 \ket{0}_2$, and the coherent state $\ket{\alpha,[\phi]} = \ket{\alpha,[\phi]}_1 \ket{0}_2$. Here the subscripts $1$ and $2$ label the two ports of the beam splitter. By definition, the vacuum state is unchanged across the BS, but the other two input states are converted to the corresponding output states by use of the Hermitian conjugate of Eqs. \eqref{s330inv}.

\subsubsection{Conversion of the single-photon state}

The input single-photon state is
\begin{align}\label{q10}
\ket{1[\phi]}_1 \ket{0}_2 = & \; \had_1[\phi] \ket{0} \nonumber \\[6pt]
= & \; \sum_\mu \phi_\mu \had_{1\mu}\ket{0},
\end{align}
where \eqref{s470} has been used. Substituting the conjugate of \eqref{s330invA} into \eqref{q10}, we obtain
\begin{align}\label{q20}
\ket{1[\phi]}_1 \ket{0}_2 = & \; \sum_\mu \phi_\mu \left[\tau \, \hbd_{1\mu}+ \rho \, (-1)^{m_\mu} \hbd_{2\mu} \right] \ket{0} \nonumber \\[6pt]
= & \; \tau \, \underbrace{\sum_\mu \phi_\mu \hbd_{1\mu}}_{= \, \hbd_1[\phi]}\ket{0} + \rho \, \underbrace{\sum_\mu (-1)^{m_\mu}\phi_\mu  \hbd_{2\mu}}_{= \, \hbd_2[\tilde{\phi}] } \ket{0},
\end{align}
where  \eqref{s470} has been used again, and we have defined
\begin{align}\label{q30}
\hbd_2[\tilde{\phi}] = & \; \sum_\mu (-1)^{m_\mu}\phi_\mu  \, \hbd_{2\mu} \nonumber \\[6pt]
= & \; \sum_\mu \hbd_{2\mu} (-1)^{m_\mu} \int \di^2 x \, u_\mu^*(x,y,z_2)\phi(x,y,z_2)  \nonumber \\[6pt]
= & \; \sum_\mu \hbd_{2\mu} \int \di^2 x \, \underbrace{(-1)^{m_\mu} u_\mu^*(x,y,z_2)}_{= \, u_\mu^*(x,-y,z_2)} \phi(x,y,z_2)  \nonumber \\
%
%= & \; \sum_\mu \int \di^2 x \,u_\mu^*(x,-y,z_2)\phi(x,y,z_2) \, \hbd_{2\mu} \nonumber \\[6pt]
%
= & \; \sum_\mu \hbd_{2\mu} \underbrace{\int \di^2 x \,u_\mu^*(x,y,z_2)\phi(x,-y,z_2)}_{ \equiv \, \tilde{\phi}_\mu} \nonumber \\
= & \; \sum_\mu \hbd_{2\mu}\tilde{\phi}_\mu  ,
\end{align}
where \eqref{s300} has been used and $\tilde{\phi}(x,y,z) \equiv \phi(x,-y,z)$. Therefore, we can rewrite \eqref{g20} as
\begin{align}\label{q40}
\ket{1[\phi]}_1 \ket{0}_2 = & \; \tau \,  \hbd_1[\phi] \ket{0} + \rho \,   \hbd_2[\tilde{\phi}] \ket{0} \nonumber \\[6pt]
= & \; \tau \ket{1[\phi]}_1 \ket{0}_2 + \rho \ket{0}_1\ket{1[\tilde{\phi}]}_2.
\end{align}

By definition of independent modes, it follows that $\bra{\Psi_1, \Psi_2, t} = \prescript{}{1}{\bra{\Psi_1,  t} } \prescript{}{2}{\bra{\Psi_2, t}}$, % = \bra{\Psi_1,  t} \bra{\Psi_2, t} $ (in the remainder we will omit the subscripts in the bras, this should not create confusion),
so that we can write
%$ \prescript{}{2}{\mathbf{C}} $
%$ \prescript{14}{2}{\mathbf{C}} $
%
\begin{align}\label{q50}
\brak{\Psi_1, \Psi_2, t}{1[\phi]} = & \; \tau \, \brak{\Psi_1,t}{1[\phi]} \brak{\Psi_2,t}{0} + \rho \, \brak{\Psi_1,t}{0} \brak{\Psi_2,t}{1[\phi]}  \nonumber \\[6pt]
= & \; \tau \, \frac{e^{-(\psi_1, \psi_1)/2}}{Z_1^{1/2}} \, \bigl( \psi_1 , \phi \bigr) e^{- i \omega t} \frac{e^{-(\psi_2, \psi_2)/2}}{Z_2^{1/2}} + \rho \, \frac{e^{-(\psi_1, \psi_1)/2}}{Z_1^{1/2}} \, \frac{e^{-(\psi_2, \psi_2)/2}}{Z_2^{1/2}} \bigl( \psi_2 , \tilde{\phi} \bigr) e^{- i \omega t}\nonumber \\[6pt]
= & \; \frac{e^{-[(\psi_1, \psi_1) + (\psi_2, \psi_2) ]/2}}{Z_1^{1/2} Z_2^{1/2}} \,  \left\{ \tau \, \bigl( \psi_1 , \phi \bigr) + \rho \, \bigl( \psi_2 , \tilde{\phi} \bigr) \right\}e^{- i \omega t}\nonumber \\[6pt]
= & \; \brak{\Psi_1, \Psi_2, t}{0}  \left\{ \tau \, \bigl( \psi_1 , \phi \bigr) + \rho \, \bigl( \psi_2 , \tilde{\phi} \bigr) \right\} e^{- i \omega t},
\end{align}
where
\begin{align}\label{q52}
Z_1^{1/2} = Z^{1/2}[\Psi_1], \qquad \text{and} \qquad Z_2^{1/2} = Z^{1/2}[\Psi_2],
\end{align}
and Eqs. \eqref{s1140} and \eqref{g80} (with $N=0$ and $N = 1$), have been used.
 %and the symbol $t$ (time) in the exponentials should not be confused with the same symbol indicating the transmission % coefficient of the beam splitter.
 For a ${50 \! : \! 50}$ beam splitter we can substitute \eqref{s340} into \eqref{q50} to obtain
\begin{align}\label{q50bis}
\brak{\Psi_1, \Psi_2, t}{1[\phi]} =  \brak{\Psi_1, \Psi_2, t}{0} \, \frac{e^{- i \omega t}}{\sqrt{2}} \left\{\bigl( \psi_1 , \phi \bigr) + i \, \bigl( \psi_2 , \tilde{\phi} \bigr) \right\} .
\end{align}

\subsubsection{Conversion of the coherent state}

The input coherent state is
\begin{align}\label{q60}
\ket{\alpha,[\phi]}_1 \ket{0}_2 = & \; \exp \left\{ \alpha \had_1[\phi] - \alpha^* \ha_1[\phi] \right\}\ket{0} \nonumber \\[6pt]
= & \; \exp \left\{ \alpha \sum_\mu \phi_\mu \had_{1\mu}  -  \alpha^* \sum_\mu \phi_\mu^* \ha_{1\mu}\right\} \ket{0},
\end{align}
where Eqs. \eqref{s470} and \eqref{s570} have been used. Substituting  Eq. \eqref{s330invA} and its conjugate into \eqref{q60}, we obtain
\begin{align}\label{q70}
\ket{\alpha,[\phi]}_1 \ket{0}_2 = & \; \exp \left\{ \alpha \sum_\mu \phi_\mu \left[\tau \, \hbd_{1\mu}+ \rho \, (-1)^{m_\mu} \hbd_{2\mu} \right]  -  \alpha^* \sum_\mu \phi_\mu^* \left[\tau^* \, \hb_{1\mu}+ \rho^* \, (-1)^{m_\mu} \hb_{2\mu} \right] \right\} \ket{0} \nonumber \\[6pt]
= & \; \exp \left\{ \sum_\mu \left[ \bigl( \tau \, \alpha \, \phi_\mu \bigr) \hbd_{1 \mu} - \bigl( \tau \, \alpha \, \phi_\mu \bigr)^* \hb_{1 \mu} \right] +  \sum_\mu \left[ \bigl( \rho \, \alpha (-1)^{m_\mu} \phi_\mu \bigr) \hbd_{2 \mu} - \bigl( \rho \, \alpha (-1)^{m_\mu} \phi_\mu \bigr)^* \hb_{2 \mu} \right] \right\} \ket{0}\nonumber \\[6pt]
= & \; \exp \left\{ \sum_\mu \left[ \bigl( \tau \, \alpha \, \phi_\mu \bigr) \hbd_{1 \mu} - \bigl( \tau \, \alpha \, \phi_\mu \bigr)^* \hb_{1 \mu} \right] \right\} \ket{0}_1   \exp \left\{\sum_\mu \left[ \bigl( \rho \, \alpha (-1)^{m_\mu} \phi_\mu \bigr) \hbd_{2 \mu} - \bigl( \rho \, \alpha (-1)^{m_\mu} \phi_\mu \bigr)^* \hb_{2 \mu} \right] \right\} \ket{0}_2,
%\nonumber \\[6pt]
%
%= & \; e^{ \sum_\mu \left[ \left( t \alpha \phi_\mu \right) \hbd_{1 \mu} - \left( t \alpha \phi_\mu \right)^* \hb_{1 \mu} \right] } \ket{0} \, e^{\sum_\mu \left[ \left( r \alpha (-1)^{m_\mu} \phi_\mu \right) \hbd_{2 \mu} - \left( r \alpha (-1)^{m_\mu} \phi_\mu \right)^* \hb_{2 \mu} \right] } \ket{0},
%
\end{align}
where  the commutation relations \eqref{s370} and \eqref{s380}, have been used. Using Eqs.  \eqref{s470},  \eqref{s570} and \eqref{q30}, we can rewrite \eqref{q70} as
\begin{align}\label{q80}
\ket{\alpha,[\phi]}_1 \ket{0}_2 = & \; \exp \left\{  \left( \tau \, \alpha  \right) \hbd_{1}[\phi] - \left( \tau \, \alpha \right)^* \hb_{1}[\phi] \right\} \ket{0}_1   \exp \left\{  \left( \rho \, \alpha  \right) \hbd_{2}[\tilde{\phi}] -  \left( \rho \, \alpha  \right)^* \hb_{2}[\tilde{\phi}] \right\} \ket{0}_2 \nonumber \\[6pt]
= & \;  \ket{\tau \, \alpha,[\phi]}_1 \ket{\rho \, \alpha,[\tilde{\phi}]}_2 ,
\end{align}
so that
\begin{align}\label{q90}
\brak{\Psi_1, \Psi_2, t}{\alpha,[\phi]} = & \; \brak{\Psi_1,t}{\tau \, \alpha,[\phi]}_1 \brak{\Psi_2,t}{\rho \, \alpha,[\tilde{\phi}]}_2 \nonumber \\[6pt]
= & \; \brak{\Psi_1,  t}{0} \, e^{-\abs{\tau \alpha}^2/2} \exp \left\{ (\tau \, \alpha) \, e^{-i \omega t} \bigl[ \psi_1 \,\phi \bigr] -\frac{1}{2} (\tau \, \alpha)^2 \, e^{-2i \omega t}  \bigl[ \phi^2 \bigr]  \right\} \nonumber \\[6pt]
 & \times  \brak{\Psi_2, t}{0} \, e^{-\abs{\rho \, \alpha}^2/2} \exp \left\{ (\rho \, \alpha) \, e^{-i \omega t}\bigl[ \psi_1 \,\tilde{\phi} \bigr] -\frac{1}{2} (\rho \, \alpha)^2 \, e^{-2i \omega t}  \bigl[ \tilde{\phi}^2 \bigr]  \right\}  \nonumber \\[6pt]
= & \; \brak{\Psi_1, \Psi_2, t}{0} \,  e^{-\left(\abs{\tau \, \alpha}^2 + \abs{\rho \, \alpha}^2 \right)/2}
 \exp \left\{e^{-i \omega t}\left[ (\tau \, \alpha) \, \bigl[ \psi_1 \,\phi \bigr] + (\rho \, \alpha) \, \bigl[ \psi_2 \, \tilde{\phi} \bigr]  \right] \right\}\nonumber \\[6pt]
 & \times \exp \left\{-  e^{-2i \omega t} \, \frac{\left(\tau \, \alpha \right)^2 + \left(\rho \, \alpha \right)^2}{2} \, \bigl[ \phi^2 \bigr]   \right\},
\end{align}
where  $\bigl[ \tilde{\phi}^2 \bigr]  = \bigl[ \phi^2 \bigr] $, and \eqref{g20} have been used. For a ${50 \! : \! 50}$ beam splitter Eqs. \eqref{s340} imply  $\left(t \alpha \right)^2 + \left(r \alpha \right)^2 = 0$, so that we can substitute \eqref{s340} into \eqref{q90} to obtain
\begin{align}\label{q100}
\brak{\Psi_1, \Psi_2, t}{\alpha,[\phi]}
= & \; \brak{\Psi_1, \Psi_2, t}{0} \,  e^{-\abs{\alpha}^2/2}
 \exp \left\{\alpha \, \frac{e^{-i \omega t} }{\sqrt{2}} \left(  \bigl[ \psi_1 \,\phi \bigr] + i \, \bigl[ \psi_2 \, \tilde{\phi} \bigr]  \right) \right\}.
\end{align}
Comparing Eq. \eqref{q50bis} with Eq. \eqref{q100} we can see that for a ${50 \! : \! 50}$ beam splitter  we obtain the particularly simple and suggestive result
\begin{align}\label{q110}
\frac{\brak{\Psi_1, \Psi_2, t}{\alpha,[\phi]}}{ \brak{\Psi_1, \Psi_2, t}{0}}
= & \; \,  e^{-\abs{\alpha}^2/2}
 \exp \left\{\alpha \, \frac{\brak{\Psi_1, \Psi_2, t}{1[\phi]}}{ \brak{\Psi_1, \Psi_2, t}{0} }  \right\}.
\end{align}

\section{Normalization of the eigenstates of the fields}

In this section we will work in the Schr\"{o}dinger picture, so that all the operators and their eigenstates will be time-independent. We want to determine the normalization terms $Z^{1/2}[\Psi]$ and $Z^{1/2}[\Pi]$ introduced in Eqs. \eqref{s1130} and \eqref{s1130bis}, respectively. To this end, we must use the functional representation of the field operators, that is \cite{Jackiw1989b},
\begin{subequations}\label{n10}
\begin{align}
\mean{\Psi}{\hPi(\mx,z)}{\Phi}  = & \; \frac{1}{i} \, \frac{\delta}{\delta \Psi(\mx,z)} \brak{\Psi}{\Phi},  \label{n10A} \\[6pt]
\mean{\Pi}{\hPsi(\mx,z)}{\Phi}  = & \; i \, \frac{\delta}{\delta \Pi(\mx,z)} \brak{\Pi}{\Phi}, \label{n10B}
\end{align}
\end{subequations}
where $\ket{\Phi}$ is a generic state vector. Replacing $\ket{\Phi}$  with $\ket{\Pi}$ in \eqref{n10A} and  with $\ket{\Psi}$ in \eqref{n10B}, we find
\begin{subequations}\label{n20}
\begin{align}
\Pi(\mx,z) \brak{\Psi}{\Pi}  = & \; \frac{1}{i} \, \frac{\delta}{\delta \Psi(\mx,z)} \brak{\Psi}{\Pi},  \label{n20A} \\[6pt]
\Psi(\mx,z) \brak{\Pi}{\Psi}  = & \; i \, \frac{\delta}{\delta \Pi(\mx,z)} \brak{\Pi}{\Psi}, \label{n20B}
\end{align}
\end{subequations}
where Eqs. \eqref{s730} and \eqref{s730bis} have been used. To satisfy \eqref{n20A} we can take
\begin{align}\label{n30}
\brak{\Psi}{\Pi} = f[\Pi]\exp \left\{ i  \int \di^2 x \, \Psi(\mx,z) \Pi(\mx,z) \right\},
\end{align}
where $f[\Pi]$ is an arbitrary functional of $\Pi(\mx,z)$. Likewise, we can satisfy \eqref{n20B} by choosing
\begin{align}\label{n40}
\brak{\Pi}{\Psi} = g^*[\Psi]\exp \left\{ -i  \int \di^2 x \, \Psi(\mx,z) \Pi(\mx,z) \right\},
\end{align}
where $g[\Psi]$ is an arbitrary functional of $\Psi(\mx,z)$. However,  $\brak{\Pi}{\Psi} = \left( \brak{\Psi}{\Pi} \right)^*$, so that we must choose
\begin{align}\label{n50}
f[\Pi] =  g[\Psi] = \text{const.}
\end{align}
Therefore, we put
\begin{align}\label{n60}
\brak{\Psi}{\Pi} =  \left( \text{const.} \right) \times \exp \left\{ i  \int \di^2 x \, \Psi(\mx,z) \Pi(\mx,z) \right\}.
\end{align}
Conventionally, we choose such a constant equal to $1$ (see, e.g., sec. 14.2.3 of \cite{SchwartzBook}).

To determine the normalization terms $Z^{1/2}[\Psi]$ and $Z^{1/2}[\Pi]$, we now require that the amplitudes $\brak{\Psi}{\Pi}$ and $\brak{\Pi}{\Psi} = \left( \brak{\Psi}{\Pi} \right)^*$, when  written  in terms of Eqs. \eqref{s1130}, would  satisfy Eqs. \eqref{n20}. First, without loss of generality let us write $1/Z^{1/2}[\Psi] = \exp \left( F[\Psi]\right)/Z^{1/2}_\Psi$, where $F[\Psi]$ is a functional of $\Psi(\mx,z)$ to be determined and $Z^{1/2}_\Psi$ now is a constant number. Next, we calculate the right-hand side of \eqref{n20A} as
\begin{align}\label{n70}
\frac{1}{i} \, \frac{\delta}{\delta \Psi(\mx,z)} \brak{\Psi}{\Pi} = & \; \frac{1}{i} \, \frac{\delta}{\delta \Psi(\mx,z)} \left\{ \frac{e^{ F[\Psi]}}{Z^{1/2}_\Psi} \, \bra{0} e^{-\frac{1}{2}  \int \di^2 x'  \left[  \hpp (\mx',z,t)  -  \sqrt{2 \omega} \, \Psi(\mx', z) \right]^2  } \ket{\Pi} \right\} \nonumber \\[6pt]
= & \; \frac{1}{i} \, \frac{\delta F[\Psi]}{\delta \Psi(\mx,z)} \brak{\Psi}{\Pi} + \frac{e^{ F[\Psi]}}{Z^{1/2}_\Psi} \, \bra{0}  \frac{1}{i} \, \frac{\delta}{\delta \Psi(\mx,z)} e^{  \int \di^2 x'  \left[  -\frac{1}{2}\hpp (\mx',z,t)^2  + \sqrt{2 \omega} \, \Psi(\mx', z) \, \hpp (\mx',z,t) -  \omega \, \Psi^2(\mx', z)   \right]  } \ket{\Pi}\nonumber \\[6pt]
= & \; \frac{1}{i} \, \frac{\delta F[\Psi]}{\delta \Psi(\mx,z)} \brak{\Psi}{\Pi}  \nonumber \\[6pt]
& + \bra{\Psi}  \frac{1}{i} \, \frac{\delta}{\delta \Psi(\mx,z)} \left\{ \int \di^2 x'  \left[  -\frac{1}{2}\hpp (\mx',z,t)^2  + \sqrt{2 \omega} \, \Psi(\mx', z) \, \hpp (\mx',z,t) -  \omega \, \Psi^2(\mx', z)   \right]  \right\} \ket{\Pi}\nonumber \\[6pt]
= & \; \frac{1}{i} \, \frac{\delta F[\Psi]}{\delta \Psi(\mx,z)} \brak{\Psi}{\Pi}   + \frac{1}{i} \, \bra{\Psi}   \sqrt{2 \omega} \, \hpp (\mx,z,t) - 2 \omega \, \Psi(\mx, z)  \ket{\Pi}.
\end{align}
Now we use Eqs. \eqref{s10} and \eqref{s70} to write
\begin{align}\label{n80}
\hpp(\mx,z) = \sqrt{\frac{\omega}{2}} \, \hPsi(\mx,z) + \frac{i}{\sqrt{2 \omega}}\,\hPi(\mx,z).
\end{align}
Substituting \eqref{n80} into \eqref{n70}, we obtain
\begin{align}\label{n90}
\frac{1}{i} \, \frac{\delta}{\delta \Psi(\mx,z)} \brak{\Psi}{\Pi} = & \; \frac{1}{i} \, \frac{\delta F[\Psi]}{\delta \Psi(\mx,z)} \brak{\Psi}{\Pi}   + \frac{1}{i} \, \bra{\Psi}   \sqrt{2 \omega} \, \left[ \sqrt{\frac{\omega}{2}} \, \hPsi(\mx,z) + \frac{i}{\sqrt{2 \omega}}\,\hPi(\mx,z) \right] - 2 \omega \, \Psi(\mx, z)  \ket{\Pi}\nonumber \\[6pt]
= & \; \frac{1}{i} \left\{ \frac{\delta F[\Psi]}{\delta \Psi(\mx,z)} + \omega \, \Psi(\mx,z) + i \, \Pi(\mx,z) - 2 \omega \, \Psi(\mx, z) \right\} \brak{\Psi}{\Pi}\nonumber \\[6pt]
= & \; \Pi(\mx,z)  \brak{\Psi}{\Pi} + \frac{1}{i} \left\{ \frac{\delta F[\Psi]}{\delta \Psi(\mx,z)} - \omega \, \Psi(\mx,z) \right\} \brak{\Psi}{\Pi}.
\end{align}
Clearly, the right-hand side of \eqref{n90} is equal to $\Pi(\mx,z)  \brak{\Psi}{\Pi}$ if and only if  $F[\Psi]$ satisfies the functional differential equation
\begin{align}\label{n100}
\frac{\delta F[\Psi]}{\delta \Psi(\mx,z)} - \omega \, \Psi(\mx,z) = 0.
\end{align}
This equation is satisfied by
\begin{align}\label{n110}
F[\Psi] = \frac{\omega}{2}  \int \di^2 x \, \Psi^2(\mx,z) + \text{const.}
\end{align}
The constant term in \eqref{n110} cannot be determined, so its value is just a matter of convention, and we choose  it equal to zero. Then, we can  write
\begin{align}\label{n120}
\frac{1}{Z^{1/2}[\Psi]} = \frac{1}{Z_\Psi^{1/2}} \, \exp \left\{ \frac{\omega}{2}  \int \di^2 x \, \Psi^2(\mx,z) \right\}.
\end{align}
Finally, substituting \eqref{n120} into \eqref{s1130}, we obtain
\begin{align}\label{n130}
\ket{\Psi } = & \; \frac{1}{Z_\Psi^{1/2}} \,\exp \left\{ \int \di^2 x \, \left[ -\frac{\omega}{2}  \Psi^2 (\mx,z) + \sqrt{2 \, \omega} \,  \Psi (\mx,z) \hpm (\mx,z) -\frac{1}{2} \hpm (\mx,z)^2 \right]  \right\} \ket{0} \nonumber \\[6pt]
= & \;  \frac{1}{Z_\Psi^{1/2}} \,\exp \left\{ -\frac{\omega}{2} \bigl[ \Psi^2 \bigr] + \sqrt{2 \, \omega} \, \bigl[ \Psi \, \hpm \bigr] -\frac{1}{2} \bigl[\hpm \, \hpm \bigr] \right\} \ket{0}.
\end{align}
Note that because $\bra{0} \hpm(\mx,z) = 0$, from \eqref{n130} it follows that
\begin{align}\label{n135}
\brak{0}{\Psi } =  \frac{1}{Z_\Psi^{1/2}} \,\exp \left\{ -\frac{\omega}{2} \bigl[ \Psi^2 \bigr] \right\} .
\end{align}

For the normalization of the conjugate momentum eigenstate $\ket{\Pi}$, we can proceed in the same way as above, first writing  $1/Z^{1/2}[\Pi] = \exp \left( G[\Pi]\right)/Z^{1/2}_\Pi$, and then calculating the right-hand side of \eqref{n20B},
\begin{align}\label{n140}
i \, \frac{\delta}{\delta \Pi(\mx,z)} \brak{\Pi}{\Psi} = & \; i \, \frac{\delta}{\delta \Pi(\mx,z)} \left\{
\frac{e^{ G[\Pi]}}{Z^{1/2}_\Pi} \, \bra{0} e^{\frac{1}{2}  \int \di^2 x'  \left[  \hpp (\mx',z,t)  - i  \sqrt{\frac{2}{ \omega}}  \Pi(\mx', z) \right]^2  } \ket{\Psi} \right\} \nonumber \\[6pt]
= & \; i \, \frac{\delta G[\Pi]}{\delta \Pi(\mx,z)} \brak{\Pi}{\Psi} + \frac{e^{ G[\Pi]}}{Z^{1/2}_\Pi} \, \bra{0}  i \, \frac{\delta}{\delta \Pi(\mx,z)} e^{  \int \di^2 x'  \left[  \frac{1}{2}\hpp (\mx',z,t)^2  - i \, \sqrt{\frac{2 }{\omega}} \, \Pi(\mx', z) \, \hpp (\mx',z,t) -  \frac{1}{\omega} \, \Pi^2(\mx', z)   \right]  } \ket{\Psi}\nonumber \\[6pt]
= & \; i \, \frac{\delta G[\Pi]}{\delta \Pi(\mx,z)} \brak{\Pi}{\Psi}  \nonumber \\[6pt]
& + i \, \bra{\Pi} \,\frac{\delta}{\delta \Pi(\mx,z)} \left\{ \int \di^2 x'   \left[  \frac{1}{2}\hpp (\mx',z,t)^2  - i \, \sqrt{\frac{2 }{\omega}} \, \Pi(\mx', z) \, \hpp (\mx',z,t) -  \frac{1}{\omega} \, \Pi^2(\mx', z)   \right]  \right\} \ket{\Psi}\nonumber \\[6pt]
= & \; i \, \frac{\delta G[\Pi]}{\delta \Pi(\mx,z)} \brak{\Pi}{\Psi}   + i \, \bra{\Pi}   - i \, \sqrt{\frac{2 }{\omega}} \,  \hpp (\mx,z,t) -   \frac{2}{\omega} \, \Pi(\mx, z) \ket{\Psi}.
\end{align}
Substituting \eqref{n80} into \eqref{n140}, we obtain
\begin{align}\label{n150}
i \, \frac{\delta}{\delta \Pi(\mx,z)} \brak{\Pi}{\Psi}  = & \; i \, \frac{\delta G[\Pi]}{\delta \Pi(\mx,z)} \brak{\Pi}{\Psi}   + i \, \bra{\Pi}   - i \, \sqrt{\frac{2 }{\omega}} \left[ \sqrt{\frac{\omega}{2}} \, \hPsi(\mx,z) + \frac{i}{\sqrt{2 \omega}}\,\hPi(\mx,z) \right] -   \frac{2}{\omega} \, \Pi(\mx, z) \ket{\Psi} \nonumber \\[6pt]
= & \; i \left\{ \frac{\delta G[\Pi]}{\delta \Pi(\mx,z)}  -i \, \Psi(\mx,z) +  \frac{1}{\omega} \, \Pi(\mx, z) -  \frac{2}{\omega} \, \Pi(\mx, z) \right\} \brak{\Pi}{\Psi}\nonumber \\[6pt]
= & \; \Psi(\mx,z)  \brak{\Pi}{\Psi} + i \left\{ \frac{\delta G[\Pi]}{\delta \Pi(\mx,z)}   -  \frac{1}{\omega} \, \Pi(\mx, z) \right\} \brak{\Pi}{\Psi}.
\end{align}
Now, the right-hand side of \eqref{n150} is equal to $\Psi(\mx,z)  \brak{\Pi}{\Psi}$ if and only if  $G[\Pi]$ satisfies the functional differential equation
\begin{align}\label{n160}
\frac{\delta G[\Pi]}{\delta \Pi(\mx,z)} - \frac{1}{\omega} \, \Pi(\mx,z) = 0.
\end{align}
This equation is satisfied by
\begin{align}\label{n170}
G[\Pi] = \frac{1}{2 \, \omega}  \int \di^2 x \, \Pi^2(\mx,z) + \text{const.}
\end{align}
Choosing, as before, the constant term equal to zero, we can  write
\begin{align}\label{n180}
\frac{1}{Z^{1/2}[\Pi]} = \frac{1}{Z^{1/2}_\Pi} \, \exp \left\{ \frac{1}{2 \, \omega}  \int \di^2 x \, \Pi^2(\mx,z) \right\}.
\end{align}
Finally,
Substituting \eqref{n180} into \eqref{s1130bis}, we obtain
\begin{align}\label{n190}
\ket{\Pi } = & \; \frac{1}{Z^{1/2}_\Pi} \, \exp \left\{ \int \di^2 x  \left[ -\frac{1}{2 \, \omega}  \Psi^2 (\mx,z) + i \, \sqrt{\frac{2}{\omega} }  \,  \Pi (\mx,z) \hpm (\mx,z) + \frac{1}{2} \hpm (\mx,z)^2 \right]  \right\} \ket{0} \nonumber \\[6pt]
= & \;  \frac{1}{Z^{1/2}_\Pi} \, \exp \left\{ -\frac{1}{2 \, \omega} \bigl[ \Pi^2 \bigr] + i \, \sqrt{\frac{2}{\omega} } \, \bigl[ \Pi \, \hpm \bigr] + \frac{1}{2} \bigl[\hpm \, \hpm \bigr] \right\} \ket{0}.
\end{align}
Note again that from $\bra{0} \hpm(\mx,z) = 0$ and \eqref{n190}, it follows that
\begin{align}\label{n200}
\brak{0}{\Pi } =  \frac{1}{Z^{1/2}_\Pi} \, \exp \left\{ -\frac{1}{2 \, \omega} \bigl[ \Pi^2 \bigr] \right\} .
\end{align}

\subsection{Completeness relations}

Following \cite{SchwartzBook}, we \emph{postulate} that

\begin{subequations}\label{m10}
\begin{align}
\int {\D \Psi} \, \proj{\Psi}{\Psi} = & \; \hId,  \label{m10A} \\[6pt]
\int {\D \Pi} \, \proj{\Pi}{\Pi} = & \; \hId, \label{m10B}
\end{align}
\end{subequations}
where $\hat{I}$ is the identity operator and $\D \Psi$ and $\D \Pi$ are the functional measures.
Then, the still unknown normalization constants $Z^{1/2}_\Psi$ and $Z^{1/2}_\Pi$ can be determined, respectively,  by the conditions
\begin{align}\label{m20}
1 = & \; \brak{0}{0} \nonumber \\[6pt]
= & \;  \int {\D \Psi} \, \brak{0}{\Psi} \brak{\Psi}{0} \nonumber \\[6pt]
= & \; \frac{1}{Z_\Psi} \int {\D \Psi} \, \exp \left\{ - \omega \bigl[ \Psi^2 \bigr] \right\},
\end{align}
and
\begin{align}\label{m30}
1 = & \; \brak{0}{0} \nonumber \\[6pt]
= & \;  \int {\D \Pi} \, \brak{0}{\Pi} \brak{\Pi}{0} \nonumber \\[6pt]
= & \;  \frac{1}{Z_\Pi} \int {\D \Pi} \, \exp \left\{ - \frac{1}{\omega} \bigl[ \Pi^2 \bigr] \right\},
\end{align}
where Eqs. \eqref{n135} and \eqref{n200} have been used. Equations \eqref{m20} and \eqref{m30} imply
\begin{align}\label{m35}
Z_\Psi =  \int {\D \Psi} \, \exp \Bigl\{ - \omega \bigl[ \Psi^2 \bigr] \Bigr\}, \qquad \text{and} \qquad  Z_\Pi =  \int {\D \Pi} \, \exp \left\{ - \frac{1}{\omega} \bigl[ \Pi^2 \bigr] \right\}.
\end{align}

Having fixed the functional measures, we can now calculate the inner product of two eigenstates of the field,
\begin{align}\label{m40}
\brak{\Psi}{\Psi'} = & \;  \int {\D \Pi} \, \brak{\Psi}{\Pi} \brak{\Pi}{\Psi'} \nonumber \\[6pt]
= & \;  \int {\D \Pi} \, \exp \Bigl\{ i \left[ \Pi \, \Psi \right] \Bigr\}  \exp \Bigl\{ -i \left[ \Pi \, \Psi' \right] \Bigr\}\nonumber \\[6pt]
= & \;  \int {\D \Pi} \, \exp \Bigl\{ i \bigl[ \Pi \left( \Psi - \Psi' \right) \bigr] \Bigr\}  ,
\end{align}
where \eqref{n60} has been used. The quantity $\brak{\Psi}{\Psi'}$ is the field-theory analogue of the quantum mechanics $\brak{q}{q'} = \delta(q-q')$. Therefore, following \cite{SchwartzBook} we formally define a functional delta function as
\begin{align}\label{m50}
\delta \left[ \Psi - \Psi' \right] = & \; \int {\D \Pi} \, \exp \left\{ i \int \di^2 x \, \Pi(\mx,z) \bigl[ \Psi(\mx,z) - \Psi'(\mx,z) \bigr] \right\} \nonumber \\[6pt]
= & \;  \int {\D \Pi} \, \exp \Bigl\{ i \bigl[ \Pi \left( \Psi - \Psi' \right) \bigr] \Bigr\}  .
\end{align}

Likewise, we calculate

\begin{align}\label{m60}
\brak{\Pi}{\Pi'} = & \;  \int {\D \Psi} \, \brak{\Pi}{\Psi} \brak{\Psi}{\Pi'} \nonumber \\[6pt]
= & \;  \int {\D \Psi} \, \exp \Bigl\{ -i \left[ \Psi \, \Pi \right] \Bigr\}  \exp \Bigl\{ i \left[ \Psi \, \Pi' \right] \Bigr\}\nonumber \\[6pt]
= & \;  \int {\D \Psi} \, \exp \Bigl\{ -i \bigl[ \Psi \left( \Pi - \Pi' \right) \bigr] \Bigr\}  ,
\end{align}
where \eqref{n60} has been used again. The quantity $\brak{\Pi}{\Pi'}$ is the field-theory analogue of  $\brak{p}{p'} = \delta(p-p')$. Then, as above, we define
\begin{align}\label{m70}
\delta \left[ \Pi - \Pi' \right] = & \; \int {\D \Psi} \, \exp \left\{ -i \int \di^2 x \, \Psi(\mx,z) \bigl[ \Pi(\mx,z) - \Pi'(\mx,z) \bigr] \right\} \nonumber \\[6pt]
= & \;  \int {\D \Psi} \, \exp \Bigl\{ -i \bigl[ \Psi \left( \Pi - \Pi' \right) \bigr] \Bigr\}  .
\end{align}

That $\brak{\Psi}{\Psi'}$ and $\brak{\Pi}{\Pi'}$ must be interpreted as delta functions, follows again from the normalization condition $\brak{0}{0} = 1$, that is
\begin{align}\label{m80}
1 = & \; \brak{0}{0} \nonumber \\[6pt]
= & \;  \int {\D \Psi} \, \brak{0}{\Psi} \brak{\Psi}{0} \nonumber \\[6pt]
= & \;  \int {\D \Psi} \,  {\D \Pi} \, {\D \Pi'} \, \brak{0}{\Pi} \brak{\Pi}{\Psi} \brak{\Psi}{\Pi'} \brak{\Pi'}{0}
\nonumber \\[6pt]
= & \;  \int {\D \Psi} \,  {\D \Pi} \, {\D \Pi'} \, \frac{\exp \Bigl\{ -\frac{1}{2 \, \omega} \bigl[ \Pi^2 \bigr] \Bigr\}}{{Z^{1/2}_{\Pi}}} \exp \Bigl\{ -i \left[ \Psi \, \Pi \right] \Bigr\}  \exp \Bigl\{ i \left[ \Psi \, \Pi' \right] \Bigr\} \frac{\exp \Bigl\{ -\frac{1}{2 \, \omega} \bigl[ {\Pi'}^2 \bigr] \Bigr\}}{{Z^{1/2}_{\Pi}}}\nonumber \\[6pt]
= & \;   \frac{1}{Z_{\Pi}}  \int {\D \Pi} \, {\D \Pi'} \left( \exp \Bigl\{ -\frac{1}{2 \, \omega} \left( \bigl[ \Pi^2 \bigr] + \bigl[ {\Pi'}^2 \bigr] \right)  \Bigr\} \underbrace{ \int {\D \Psi} \, \exp \Bigl\{ -i \bigl[ \Psi \left( \Pi - \Pi' \right) \bigr] \Bigr\} }_{= \, \delta [\Pi - \Pi' ]} \right)\nonumber \\[6pt]
= & \; \frac{1}{Z_{\Pi}} \int {\D \Pi}  \exp \Bigl\{ -\frac{1}{ \omega} \bigl[ \Pi^2 \bigr] \Bigr\},
\end{align}
which coincides with \eqref{m30}. A very similar calculation can be easily done to show that $\delta [\Psi - \Psi' ]$ defined by \eqref{m50}, must be interpreted as a functional delta.

%\bibliography{all_bibliography_2021}